\documentclass[10pt,twocolumn,letterpaper]{article}

\usepackage[pagenumbers]{cvpr}

\usepackage{microtype}
\usepackage{multirow}
\usepackage{tabularx}
\usepackage{flafter}

\definecolor{revpurple}{RGB}{112,48,160}

\newcommand{\artsim}{ArtSim}
\newcommand{\ours}{\textsc{Patina}}
\renewcommand{\topfraction}{0.85}
\renewcommand{\dbltopfraction}{0.85}

\definecolor{cvprblue}{rgb}{0.21,0.49,0.74}
\usepackage[pagebackref,breaklinks,colorlinks,allcolors=cvprblue]{hyperref}
\AtEndPreamble{
  \crefname{figure}{Figure}{Figures}
  \Crefname{figure}{Figure}{Figures}
  \crefformat{section}{\S#2#1#3}
  \Crefformat{section}{\S#2#1#3}
  \crefformat{subsection}{\S#2#1#3}
  \Crefformat{subsection}{\S#2#1#3}
  \crefmultiformat{section}{\S#2#1#3}{ and~#2#1#3}{, #2#1#3}{, and~#2#1#3}
  \Crefmultiformat{section}{\S#2#1#3}{ and~#2#1#3}{, #2#1#3}{, and~#2#1#3}
  \crefformat{appendix}{\S#2#1#3}
  \Crefformat{appendix}{\S#2#1#3}
  \crefformat{subappendix}{\S#2#1#3}
  \Crefformat{subappendix}{\S#2#1#3}
  \crefmultiformat{appendix}{\S#2#1#3}{ and~#2#1#3}{, #2#1#3}{, and~#2#1#3}
  \Crefmultiformat{appendix}{\S#2#1#3}{ and~#2#1#3}{, #2#1#3}{, and~#2#1#3}
  \crefname{table}{Table}{Tables}
  \Crefname{table}{Table}{Tables}
  \crefformat{equation}{Eq.~#2#1#3}
  \Crefformat{equation}{Eq.~#2#1#3}
  \crefmultiformat{equation}{Eqs.~#2#1#3}{ and~#2#1#3}{, #2#1#3}{, and~#2#1#3}
  \Crefmultiformat{equation}{Eqs.~#2#1#3}{ and~#2#1#3}{, #2#1#3}{, and~#2#1#3}
}

\usepackage[most]{tcolorbox}
\newtcolorbox{takeaway}{enhanced, colback=cvprblue!7, colframe=cvprblue!80, boxrule=0.6pt, arc=2pt,
  left=5pt, right=5pt, top=2pt, bottom=2pt, fontupper=\small, before skip=8pt, after skip=8pt}
\newlength{\qbadgedp}\newlength{\qbadgeht}
\newcommand{\qbadge}[1]{\setlength{\qbadgedp}{0.75ex}\setlength{\qbadgeht}{2.5ex}%
  \tcbox[on line, colback=cvprblue, colframe=cvprblue, coltext=white, boxrule=0pt, arc=2pt,
  boxsep=0pt, left=2.5pt, right=2.5pt, top=0pt, bottom=0pt, fontupper=\bfseries\footnotesize]{\rule[-\qbadgedp]{0pt}{\qbadgeht}Q#1}}

\usepackage{soul}
\newcommand{\qhl}[1]{{\sethlcolor{cvprblue!20}\hl{#1}}}

\def\confName{CVPR}
\def\confYear{2027}

\title{How Well Is My Artwork Protected?\\Predicting and Tailoring Protection against Style Mimicry by Personalization}

\author{Jihye Kang\thanks{Equal contribution.} \qquad Jiwon Park\footnotemark[1] \qquad Sungwoo Hwang \qquad Namhyuk Ahn\\
Inha University}

\begin{document}
\maketitle
\vspace{-10pt}
\begin{figure}[t]
  \centering
  \includegraphics[width=\linewidth]{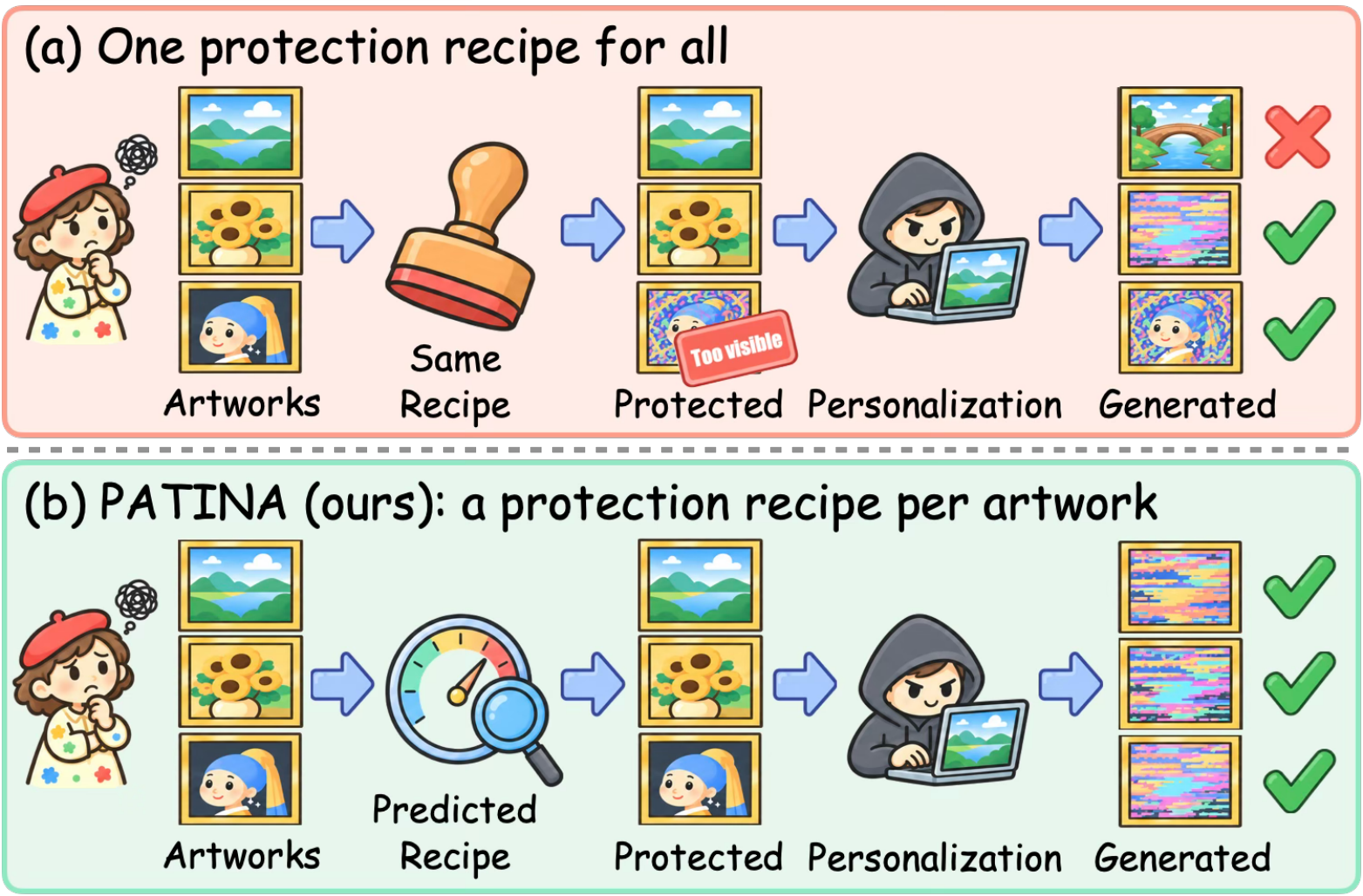}
  \par\vspace{-12pt}
  \caption{\textbf{Artwork-specific protection.}
A single recipe can leave some artworks vulnerable while
unnecessarily perturbing others.
\ours{} predicts protection before personalization and selects
the weakest recipe expected to meet a target protection level.}
  \label{fig:teaser}
  \vspace{-17pt}
\end{figure}

\begin{abstract}
Protection tools add an imperceptible perturbation to each artwork so that a model personalized on it fails to imitate its style, and they apply one recipe, a protection method at a fixed strength, to every artwork.
We find that the same recipe protects artworks very differently, well beyond the noise of personalization seeds, and that cheap signals predict this protection before any personalization.
Because protection rises and falls together across recipes, our method, \ours{}, predicts the protection of every recipe from one short trial protection and gives each artwork the weakest recipe whose predicted protection reaches a chosen level.
At the same level of visibility, \ours{} lets more artworks pass the protection threshold than any single recipe and warns the artist about artworks for which no recipe is predicted to pass it.
\end{abstract}

\section{Introduction}
\label{sec:intro}

Personalization methods~\cite{ruiz2023dreambooth,hu2022lora} adapt a text-to-image diffusion model~\cite{rombach2022ldm,podell2024sdxl} to a few images~\cite{gal2023textual,kumari2023multi}, and a single artwork is enough to reproduce its style in new images~\cite{sohn2023styledrop,ahn2024dreamstyler,frenkel2024blora,shah2024ziplora}.
Artists respond with protection tools, which add an imperceptible perturbation to an artwork before posting it~\cite{shan2023glaze,liang2023mist}.
Many protection methods optimize this perturbation so that a model personalized on the protected image fails to imitate it~\cite{liang2023advdm,vanle2023antidreambooth,liu2024metacloak,zheng2025ace,xue2024sds,li2025styleguard,styleprotect2026,ahn2024impasto,ahn2025fastprotect}.
We call a pairing of a protection method and a perturbation strength a \emph{protection recipe}.
Protection recipes are compared by their average over artworks, and a protection tool applies one recipe to every artwork (\cref{fig:teaser}a).

However, the average over artworks hides how protection varies from one artwork to another.
An artist protects and posts individual artworks, so what matters to the artist is how well each of these artworks is protected.
Prior work observes that the level of protection varies widely between artworks and artists~\cite{ahn2024impasto,honig2025adversarial}, that an artwork that is hard to protect for one method tends to be hard for the others~\cite{ahn2025fastprotect}, and that existing tools give artists no way to check how vulnerable their artworks are~\cite{honig2025adversarial}.
Under one protection recipe, most artworks thus fall into one of two cases.
Either the recipe is too weak for the artwork, so that the artist posts it with a false sense of security, or a weaker recipe already suffices, so that the artwork receives a more visible perturbation than it needs.
Telling the two cases apart requires protecting the artwork and personalizing a model on the protected image, which places a heavy burden on the artist, since personalization needs a GPU and more computation than protection.

For real artists, protection therefore needs to be analyzed for each artwork, and prior work reports the variation across artworks only as an observation.
This paper studies protection for each artwork on a large-scale benchmark of artworks and protection recipes.
The study keeps existing protection methods unchanged and asks whether a protection tool can evaluate, predict, and choose their protection recipes for each artwork.
Building such a tool requires answers to the following three questions, and we answer each of them through experiments and analysis on our benchmark.

\vspace{1.5pt}\noindent\qbadge{1}~\textbf{\qhl{Does one recipe protect all artworks equally?}}
Under the same approach, protection differs between artworks well beyond the noise of personalization, and the difference remains when the visibility of the perturbation is matched across artworks.
An artwork that is hard to protect under one method tends to be hard under the others, so the difficulty of protection is largely a property of the artwork.
The method that protects an artwork best, however, can still differ from one artwork to another.
A recipe average therefore does not describe how well a single artwork is protected (\cref{sec:variation}).

\vspace{1.5pt}\noindent\qbadge{2}~\textbf{\qhl{Can an artist know in advance how well their artwork is protected?}}
Protection can then be predicted from the artwork and its protected image before malicious personalization.
In our analysis, a tiny regressor can predict the protection score from signals of the artwork and the response of a frozen diffusion model to the protected image, nearly as accurately as one personalization run.
The regressor is trained once, and the artist then runs only frozen models at a fraction of the cost of one personalization, so the artist can check an artwork before posting it (\cref{sec:predict}).

\vspace{1.5pt}\noindent\qbadge{3}~\textbf{\qhl{Can we recommend a recipe for each artwork?}}
An artist asks which protection method works best for an artwork and how strong its perturbation needs to be to meet the level of protection they want.
Answering this requires protecting the artwork with every protection recipe and personalizing a model on each protected image.
The regressor above removes the personalization but still needs an image protected with each recipe.
Fortunately, the protection of an artwork rises and falls together across recipes, so a single short trial protection is enough to predict the protection under every method and strength (\cref{sec:variation}).

Our method, \ours{}, predicts every recipe from this trial with the regressor above, the \ours{} predictor, and recommends the weakest recipe whose predicted protection reaches a chosen level.
By giving stronger perturbations to hard artworks and weaker ones to easy artworks, it lets more artworks pass the protection threshold than any single recipe at the same visibility and reaches the same pass rate with a less visible perturbation (\cref{fig:teaser}b).
It also warns the artist in advance about the artworks for which no recipe within the artist's visibility limit is predicted to pass the protection threshold.
When an imitator applies strong purification~\cite{honig2025adversarial} that removes most of the protection, no recipe lets many artworks pass the threshold, and this early warning tells the artist before posting which artworks these are (\cref{sec:recipe}).

\section{Preliminaries and Setup}
\begin{figure*}[t]
  \centering
  \includegraphics[width=\linewidth]{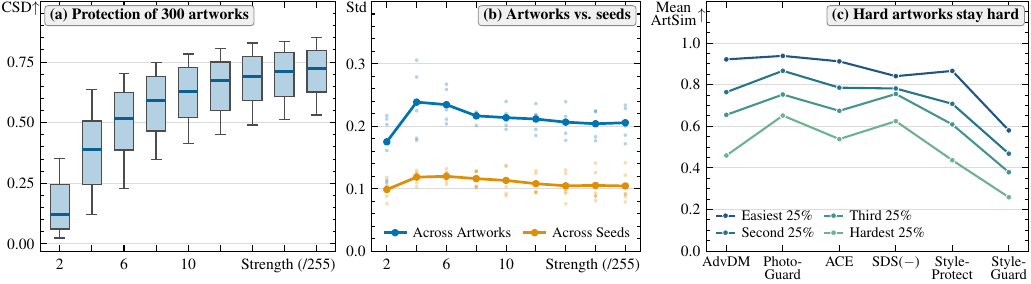}
  \par\vspace{-11pt}
  \caption{\textbf{Protection varies across artworks.} (a)~Protection score of artworks at each strength, averaged over the six protection methods. The same strength protects artworks to very different degrees. (b)~Standard deviation within one recipe across artworks (within each seed) and across seeds (within each artwork), each the square root of the mean variance. Lines average the methods and dots show each method. At every strength, artworks differ more than seeds. (c)~Mean \artsim{} protection score under each of the six methods for four groups of artworks, formed by their mean score over the other methods.}
  \label{fig:variation}
  \vspace{-12pt}
\end{figure*}

\label{sec:setup}

\subsection{Style Personalization and Protection}
\label{sec:prelim}

\smallskip\noindent\textbf{Style personalization.}
A latent diffusion model~\cite{rombach2022ldm} trains a denoiser $\epsilon_\theta$ by minimizing $\mathbb{E}_{t,\epsilon}\|\epsilon-\epsilon_\theta(z_t,t,c)\|_2^2$, where $\epsilon$ is the noise added to the latent $z_t$ of an image at step $t$ and $c$ is a text condition.
Personalization fine-tunes the denoiser or its low-rank adapters (LoRA) on a few images from a user with this loss~\cite{ruiz2023dreambooth,hu2022lora,kumari2023multi}.
Style personalization fine-tunes on artworks, and the model then imitates the style of the artist~\cite{shan2023glaze,liang2023mist,honig2025adversarial}, even from a single artwork~\cite{sohn2023styledrop,ahn2024dreamstyler,frenkel2024blora,shah2024ziplora}.
We study this single-artwork setting because it ties the outcome of protection to one artwork, so the analysis is free of the combined effect of several artworks that are personalized together in one model.

\smallskip\noindent\textbf{Image protection.}
An artist protects an artwork $x$ by posting $\tilde{x}=x+\delta$, where $\|\delta\|_\infty\le\varepsilon$ and $\delta$ is optimized against a diffusion model so that personalization on $\tilde{x}$ fails to imitate the style.
Protection methods raise~\cite{liang2023advdm,vanle2023antidreambooth,liu2024metacloak,styleprotect2026} or retarget~\cite{zheng2025ace} the denoising loss, shift the latent of the image encoder~\cite{salman2023photoguard,li2024pid}, reverse the gradient of score distillation~\cite{xue2024sds}, or target style features~\cite{shan2023glaze,li2025styleguard}.

We call the pair $r=(m,\varepsilon)$ of a protection method $m$ and a strength $\varepsilon$ a \emph{protection recipe}, as it fixes what the artist applies and how much.
Protection work evaluates each recipe by its average over artworks~\cite{ye2025evaluating}, and a protection tool applies one recipe to every artwork.
Some protections adapt the perturbation to the artwork within one method, through a target style for each artist~\cite{shan2023glaze}, a choice among pretrained perturbations~\cite{ahn2025fastprotect}, or a perceptual map and a difficulty estimate~\cite{ahn2024impasto}.
These adaptations keep the method fixed and do not show the artist how well an artwork is protected.

\subsection{Setup}
\label{sec:measure}

\smallskip\noindent\textbf{Evaluation.}
For an artwork $x$ and a recipe $r$, we personalize on the original and the protected artwork with the same seed and prompts.
We measure the generated images with CSD~\cite{somepalli2024csd} and \artsim{}~\cite{wright2022artfid} for style similarity and CLIP-IQA~\cite{wang2023clipiqa} for image quality.
We rescale each metric into a protection score $s(x,r)$, which is 0 when the protected personalization imitates the artwork as well as the original one and 1 when it does no better than the base model or, for CLIP-IQA, at zero image quality (\cref{app:score}).
An artwork passes the \emph{protection threshold} $\tau$ when $s\ge\tau$, and the \emph{pass rate} is the percentage of artworks that pass it.
In our study, five raters choose the score at which about 50\% or 80\% of the generated images no longer show the style or are too distorted to use, and the median choice gives the \emph{50\%} and \emph{80\% thresholds} for quantitative evaluation.
However, each artist can freely set a tighter or looser threshold.
The visibility of the perturbation is $\mathrm{LPIPS}(x,\tilde{x})$~\cite{zhang2018lpips}.

\smallskip\noindent\textbf{Protection recipes and data.}
We use six existing protection methods, AdvDM~\cite{liang2023advdm}, PhotoGuard~\cite{salman2023photoguard}, ACE~\cite{zheng2025ace}, SDS($-$)~\cite{xue2024sds}, StyleProtect~\cite{styleprotect2026}, and StyleGuard~\cite{li2025styleguard}, with strengths $\varepsilon$ from 2/255 to 18/255 in steps of 2/255.
Protection and personalization are one-shot on SD~2.1~\cite{rombach2022ldm}, and each artwork is protected alone and personalized alone by training LoRA~\cite{hu2022lora}.
In our benchmark, we use 300 artworks from WikiArt~\cite{saleh2015wikiart}, StyleBench~\cite{gao2024styleshot}, and OmniStyle~\cite{wang2025omnistyle}, ranging from historical paintings to modern digital art.

\section{Unequal Protection across Artworks}
\label{sec:variation}

The average score of a recipe is useful to an artist only if every artwork is protected about equally well under it.
In our benchmark, however, the protection score at the same strength spreads widely across artworks (\cref{fig:variation}a).
Under the same recipe, personalization on the protected images fails to imitate some artworks and still imitates others (\cref{fig:qual-compare}).
One can argue that this spread is due to the randomness of personalization.
\Cref{fig:variation}b shows otherwise, as the spread across artworks exceeds that across seeds in every recipe.
Comparing the artworks at the same LPIPS leaves most of this spread, so it does not come from differences in visibility (\cref{tab:matched}).
The spread also remains when the imitator personalizes another model, fine-tunes the whole denoiser, or trains on several artworks of one artist together (\cref{app:backbones}, \cref{tab:sets-spread}).
How well a recipe protects an artwork is therefore a property of the artwork, and an artist who trusts the recipe average may post an artwork below the protection threshold.

\begin{table}[t]
  \centering
  \caption{\textbf{Predicting protection before personalization.} The middle block adds signal categories, and the \ours{} predictor uses the first three. False pass is the percentage of artworks predicted to pass the 50\% threshold ($\hat{s}\ge\tau$) whose measured score is below it ($s<\tau$). Cost is relative to one personalization. IQA is CLIP-IQA. Bold marks the best predictor besides direct personalization.}
  \label{tab:predict}
  \footnotesize
  \setlength{\tabcolsep}{0pt}
  \begin{tabular*}{\linewidth}{@{\extracolsep{\fill}}lccccccc@{}}
    \toprule
    & & \multicolumn{3}{c}{RMSE} & \multicolumn{3}{c}{False Pass (\%)} \\
    \cmidrule{3-5} \cmidrule{6-8}
    Predictor & Cost & CSD\,$\downarrow$ & \artsim{}\,$\downarrow$ & IQA\,$\downarrow$ & CSD\,$\downarrow$ & \artsim{}\,$\downarrow$ & IQA\,$\downarrow$ \\
    \midrule
    Mean Report & 0.00 & 0.213 & 0.241 & 0.160 & 19.2 & 17.3 & 19.3 \\
    DAP~\cite{ahn2024impasto} & $<$0.01 & 0.191 & 0.209 & 0.147 & 15.2 & 13.5 & 16.4 \\
    Image Embedding & $<$0.01 & 0.189 & 0.216 & 0.148 & 16.0 & 14.5 & 16.2 \\
    \midrule
    Image Signals & 0.04 & 0.186 & 0.212 & 0.146 & 15.0 & 13.9 & 16.8 \\
    + Protection Trace & 0.04 & 0.181 & 0.202 & 0.142 & 14.3 & 12.7 & 15.5 \\
    + Protected Image & 0.12 & \textbf{0.166} & 0.198 & \textbf{0.139} & \textbf{12.8} & \textbf{12.3} & 15.1 \\
    + LoRA Probe & 0.27 & \textbf{0.166} & \textbf{0.197} & \textbf{0.139} & 12.9 & 12.4 & \textbf{15.0} \\
    \midrule
    Direct Personalization & 1.00 & 0.153 & 0.189 & 0.143 & 9.5 & 9.5 & 11.7 \\
    \bottomrule
  \end{tabular*}
\end{table}

The difficulty of an artwork is also largely shared across recipes.
Using \artsim{}, we split the artworks into four equal groups by their mean protection score over the other methods and compare the groups under each method (\cref{fig:variation}c).
Every method keeps the order of the groups, so an artwork that is hard for some methods tends to be hard for the others.
This shared structure lets a few measurements of an artwork predict its protection across all methods and strengths, which is the principle of \ours{}.
The method that protects an artwork best can still differ from one artwork to another, so \ours{} chooses the method together with the strength.

\begin{takeaway}
\qbadge{1}~\textbf{Takeaway.} The average of a recipe does not tell an artist how well their own artwork is protected, so protection has to be checked for each artwork.
Since the difficulty of an artwork is largely shared across recipes, a few measurements of the artwork can predict its protection under every recipe.
\end{takeaway}

\section{Predicting Protection in Advance}
\label{sec:predict}

Checking how well an artwork is protected requires personalizing a model on the protected image, which costs a GPU and more computation than the protection itself.
However, protection is largely a property of the artwork (\cref{sec:variation}), so we expect it to be predictable from the artwork and its protected image without any personalization.
To test this, we predict the protection score $s(x,r)$ of an artwork $x$ under a recipe $r$ and compare it with an actual personalization run.

\subsection{Signals and Predictor}
\label{sec:signals}

\smallskip\noindent\textbf{Signals.}
We start from 131 candidate signals and keep 63 of them using only their correlations with one another (\cref{app:signals}).
We group the signals into four categories by when they become available and what they cost (\cref{tab:predict}).
\emph{Image} signals come from the artwork before protection and describe its frequency content, texture, color, and the responses of the VAE and the base model, at about 4\% of the cost of one personalization.
\emph{Protection trace} signals are recorded during the protection optimization, namely the loss curve, the input gradients, and the magnitude, frequency, and location of the perturbation, and they come at no extra cost.
\begin{table}[t]
  \centering
  \caption{\textbf{Prediction for unseen sources and protection methods.} Values are RMSE. Each column group leaves one source or method out of training, and for an unseen method every prediction is calibrated with five personalized artworks (\cref{app:unseen}). IQA is CLIP-IQA. Bold marks the best predictor besides direct personalization.}
  \label{tab:general}
  \vspace{1.5pt}
  \footnotesize
  \setlength{\tabcolsep}{0pt}
  \begin{tabular*}{\linewidth}{@{\extracolsep{\fill}}lcccccc@{}}
    \toprule
    & \multicolumn{3}{c}{Unseen Source} & \multicolumn{3}{c}{Unseen Method} \\
    \cmidrule{2-4} \cmidrule{5-7}
    Predictor & CSD\,$\downarrow$ & \artsim{}\,$\downarrow$ & IQA\,$\downarrow$ & CSD\,$\downarrow$ & \artsim{}\,$\downarrow$ & IQA\,$\downarrow$ \\
    \midrule
    Mean Report & 0.227 & 0.260 & 0.163 & 0.232 & 0.262 & 0.174 \\
    DAP~\cite{ahn2024impasto} & 0.199 & 0.214 & 0.146 & 0.211 & 0.231 & 0.162 \\
    Image Embedding & 0.199 & 0.230 & 0.149 & 0.215 & 0.238 & 0.165 \\
    \textbf{\ours{} Predictor} & \textbf{0.172} & \textbf{0.198} & \textbf{0.141} & \textbf{0.194} & \textbf{0.225} & \textbf{0.161} \\
    Direct Personalization & 0.152 & 0.176 & 0.134 & 0.153 & 0.189 & 0.143 \\
    \bottomrule
  \end{tabular*}
\end{table}

The other two categories measure how models respond to the finished protected image.
\emph{Protected image} signals are the responses of frozen models, such as the VAE and the denoiser, and of SDEdit~\cite{meng2022sdedit}, at about 7.5\% of the cost.
Personalization minimizes the denoising loss on noised versions of the protected image (\cref{sec:prelim}), and SDEdit on the protected image previews this process without training.
A perturbation that pushes the output of SDEdit away from the style of the artwork therefore also disrupts what personalization learns.
In our experiments, SDEdit gives one of the most predictive signals of protection (\cref{app:signals}).
\emph{LoRA probe} signals approximate personalization more closely with one to five steps of LoRA training on the protected image, at about 15\% of the cost of one personalization.

\smallskip\noindent\textbf{Predictor and baselines.}
For each protection method, one LightGBM regressor~\cite{ke2017lightgbm} predicts the protection scores of all nine strengths, with a monotone increasing constraint on strength.
We call it the \ours{} predictor.
We compare it with four baselines.
\emph{Mean report} assigns every artwork the average score of the recipe over the training artworks.
This is the information that recipe averages give an artist.
\emph{DAP} feeds the difficulty estimate of Impasto~\cite{ahn2024impasto}, a single LPIPS value computed on the protected image of the recipe, and the strength to the same regressor.
\emph{Image embedding} predicts the scores of all recipes from a generic description of the artwork, its DINOv2 embedding~\cite{oquab2024dinov2}, with a ridge regression.
\emph{Direct personalization} personalizes a model once on the protected image and uses the protection score of this run as its prediction for the artwork, and it costs one full personalization per recipe.
The metric is the root-mean-square error (RMSE) across artworks, computed for each method, strength, and seed and then averaged.

\begin{table}[t]
  \centering
  \caption{\textbf{Prediction for other personalization settings.} Values are RMSE. All predictors are trained with LoRA on SD~2.1 and applied ($\rightarrow$) to DreamBooth~\cite{ruiz2023dreambooth} on SD~2.1 or to LoRA on SD~1.5 and SDXL. Each value averages AdvDM, PhotoGuard, and StyleGuard at 16/255, and each prediction, including mean report, is calibrated on the other artworks personalized in the new setting.}
  \label{tab:transfer}
  \footnotesize
  \setlength{\tabcolsep}{0pt}
  \begin{tabular*}{\linewidth}{@{\extracolsep{\fill}}lcccccc@{}}
    \toprule
    & \multicolumn{2}{c}{$\rightarrow$\,DreamBooth} & \multicolumn{2}{c}{$\rightarrow$\,SD~1.5} & \multicolumn{2}{c}{$\rightarrow$\,SDXL} \\
    \cmidrule{2-3} \cmidrule{4-5} \cmidrule{6-7}
    Predictor & CSD\,$\downarrow$ & \artsim{}\,$\downarrow$ & CSD\,$\downarrow$ & \artsim{}\,$\downarrow$ & CSD\,$\downarrow$ & \artsim{}\,$\downarrow$ \\
    \midrule
    Mean Report & 0.231 & 0.314 & 0.247 & 0.267 & 0.211 & 0.242 \\
    DAP~\cite{ahn2024impasto} & 0.215 & 0.292 & 0.234 & 0.242 & 0.196 & 0.228 \\
    Image Embedding & 0.219 & 0.297 & 0.235 & 0.250 & 0.199 & 0.232 \\
    \textbf{\ours{} Predictor} & \textbf{0.185} & \textbf{0.285} & \textbf{0.213} & \textbf{0.235} & \textbf{0.165} & \textbf{0.226} \\
    \bottomrule
  \end{tabular*}
\end{table}

\subsection{Results}
\label{sec:predict-results}
\vspace{-1.45pt}

\begin{table*}[t]
  \centering
  \caption{\textbf{Comparison of strategies that choose a recipe for each artwork.} We compare how many artworks each strategy lets pass the protection threshold under a limit on the mean LPIPS of the protected images, and how visible its perturbations become for a 90\% pass rate. Uniform recipe follows current protection tools, and direct personalization chooses with scores measured by personalizing every recipe once with one seed. LPIPS cap protects each artwork several times to find the strongest strength within the limit, so its cost ranges from about 0.67 with a binary search over the strengths to 2.67 with all nine strengths. Each block fixes the threshold $\tau$, set where raters judge about 50\% or 80\% of the generated images to have lost the style or to be too distorted to use (\cref{sec:measure}), and the mean LPIPS limit, at which each strategy sets its knob. The first column group gives the percentage of artworks below the threshold ($s<\tau$) at this limit, the second the same percentage among the hardest 25\% of the artworks, which need the strongest recipes, and the third the mean LPIPS at which 90\% of the artworks pass. Bold marks the best strategy other than direct personalization, and a dash indicates a strategy that does not reach a 90\% pass rate.}
  \label{tab:main}
  \footnotesize
  \setlength{\tabcolsep}{0pt}
  \begin{tabular*}{\linewidth}{@{\extracolsep{\fill}}lcccccccccc@{}}
    \toprule
    \multirow{2}{*}[-2.5pt]{Strategy} & \multirow{2}{*}[-2.5pt]{Cost} & \multicolumn{3}{c}{$s<\tau$, All Artworks (\%)} & \multicolumn{3}{c}{$s<\tau$, Hardest 25\% (\%)} & \multicolumn{3}{c}{LPIPS for a 90\% Pass Rate} \\
    \cmidrule(lr){3-5}\cmidrule(lr){6-8}\cmidrule(lr){9-11}
    & & CSD\,$\downarrow$ & \artsim{}\,$\downarrow$ & CLIP-IQA\,$\downarrow$ & CSD\,$\downarrow$ & \artsim{}\,$\downarrow$ & CLIP-IQA\,$\downarrow$ & CSD\,$\downarrow$ & \artsim{}\,$\downarrow$ & CLIP-IQA\,$\downarrow$ \\
    \midrule
    \multicolumn{11}{l}{\emph{50\% threshold ($\tau$) with a mean LPIPS limit of 0.15}} \\
    Uniform Recipe & 0.00 & 17.8 & 18.4 & 10.1 & 48.4 & 49.1 & 22.4 & 0.225 & 0.179 & 0.156 \\
    Image Embedding & $<$0.01 & 6.6 & 4.0 & 5.7 & 11.8 & 10.1 & \textbf{11.0} & 0.120 & 0.120 & 0.110 \\
    DAP~\cite{ahn2024impasto} & 0.17 & 7.3 & 4.8 & 6.0 & 9.6 & 11.7 & 11.4 & 0.137 & 0.128 & 0.112 \\
    LPIPS Cap & 0.67--2.67 & 7.6 & 3.7 & 6.1 & 9.9 & 7.8 & 11.2 & 0.135 & \textbf{0.090} & 0.113 \\
    \textbf{\ours{}} & 0.28 & \textbf{3.8} & \textbf{1.7} & \textbf{4.3} & \textbf{3.0} & \textbf{4.7} & 11.1 & \textbf{0.110} & \textbf{0.090} & \textbf{0.087} \\
    Direct Personalization & 71.7 & 1.4 & 4.6 & 1.7 & 3.5 & 7.3 & 2.0 & 0.077 & 0.075 & 0.083 \\
    \midrule
    \multicolumn{11}{l}{\emph{80\% threshold ($\tau$) with a mean LPIPS limit of 0.20}} \\
    Uniform Recipe & 0.00 & 34.5 & 11.6 & 21.8 & 74.1 & 37.6 & 50.0 & -- & 0.206 & -- \\
    Image Embedding & $<$0.01 & 31.5 & 5.3 & 14.0 & 45.3 & 10.8 & 30.3 & -- & 0.161 & -- \\
    DAP~\cite{ahn2024impasto} & 0.17 & 29.3 & 6.1 & 18.3 & 47.5 & 18.3 & 37.2 & -- & 0.164 & -- \\
    LPIPS Cap & 0.67--2.67 & 31.9 & 4.8 & 16.9 & 53.5 & 11.6 & 32.8 & -- & 0.123 & -- \\
    \textbf{\ours{}} & 0.28 & \textbf{22.8} & \textbf{3.3} & \textbf{13.9} & \textbf{33.1} & \textbf{9.4} & \textbf{27.4} & \textbf{0.275} & \textbf{0.118} & \textbf{0.259} \\
    Direct Personalization & 71.7 & 9.9 & 11.7 & 11.5 & 18.0 & 14.7 & 21.3 & 0.182 & 0.110 & 0.190 \\
    \bottomrule
  \end{tabular*}
\end{table*}

\begin{figure*}[t]
  \centering
  \includegraphics[width=\linewidth]{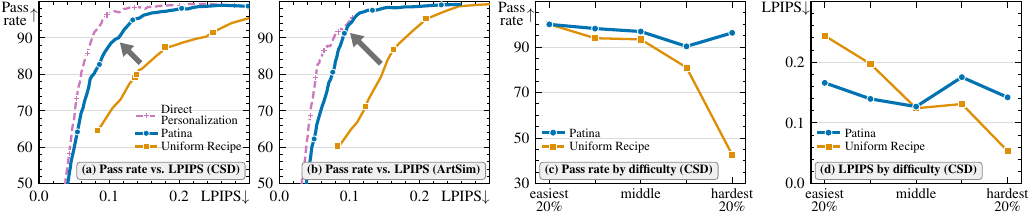}
  \par\vspace{-8pt}
  \caption{\textbf{Choosing a recipe for each artwork.} (a,~b)~Pass rate (\%), the percentage of artworks with $s\ge\tau$ (\cref{sec:measure}), against mean LPIPS. At every mean LPIPS, \ours{} lets more artworks pass the protection threshold than uniform recipe. (c,~d)~Pass rate and mean LPIPS of five groups of artworks ordered by the difficulty of protection, the weakest strength at which they pass the threshold, at mean LPIPS 0.15 under CSD. \ours{} gives the two easiest groups a less visible perturbation and the two hardest groups a stronger one, which raises their pass rate.}
  \label{fig:q3}
\end{figure*}

\smallskip\noindent\textbf{Comparison with baselines.}
Direct personalization measures protection with an actual personalization run, and its error comes only from the randomness of personalization across seeds, so it shows how accurately one personalization run per recipe checks protection.
Among the predictions that train no model, the \ours{} predictor is the most accurate on every score in \cref{tab:predict} and comes close to direct personalization with little computation.
Mean report treats all artworks of a recipe alike and has the largest error, which nearly equals the spread across artworks in \cref{fig:variation}b.
DAP~\cite{ahn2024impasto} summarizes the protected image in a single number and image embedding describes the artwork by a generic embedding, and the errors of both fall between those of mean report and the \ours{} predictor.
The false pass rate follows the same order, and a false pass is the most harmful error for an artist, who then posts an artwork below the protection threshold with a false sense of security.
An artist can thus check an artwork before posting it nearly as reliably as with one personalization run.

\smallskip\noindent\textbf{Signal ablation.}
In the middle block of \cref{tab:predict}, the image signals, measured on the artwork before protection, already match image embedding, consistent with the finding of \cref{sec:variation} that protection is largely a property of the artwork.
Adding the protection trace and then the protected image reduces the error further, since the protected image holds both the artwork and the perturbation of the recipe.
Adding the LoRA probe barely changes the error and more than doubles the cost, so the \ours{} predictor keeps only the first three categories, none of which trains a model.

\begin{table*}[t]
  \centering
  \caption{\textbf{Visibility at an equal pass rate.} We set each strategy so that 90\% of the artworks pass the 50\% threshold ($s\ge\tau$) and compare the mean $L_\infty$ strength (/255), PSNR (dB), and LPIPS of its perturbations. Per-source and per-cluster recipes choose one recipe for each source or style cluster. Bold marks the best strategy besides direct personalization, and a dash one that does not reach a 90\% pass rate.}
  \label{tab:visibility}
  \footnotesize
  \setlength{\tabcolsep}{0pt}
  \begin{tabular*}{\linewidth}{@{\extracolsep{\fill}}lcccccccccc@{}}
    \toprule
    \multirow{2}{*}[-2.5pt]{Strategy} & \multirow{2}{*}[-2.5pt]{Cost} & \multicolumn{3}{c}{CSD, $\Pr(s\ge\tau)=90\%$} & \multicolumn{3}{c}{\artsim{}, $\Pr(s\ge\tau)=90\%$} & \multicolumn{3}{c}{CLIP-IQA, $\Pr(s\ge\tau)=90\%$} \\
    \cmidrule(lr){3-5}\cmidrule(lr){6-8}\cmidrule(lr){9-11}
    & & $L_\infty$\,$\downarrow$ & PSNR\,$\uparrow$ & LPIPS\,$\downarrow$ & $L_\infty$\,$\downarrow$ & PSNR\,$\uparrow$ & LPIPS\,$\downarrow$ & $L_\infty$\,$\downarrow$ & PSNR\,$\uparrow$ & LPIPS\,$\downarrow$ \\
    \midrule
    Uniform Recipe & 0.00 & 9.7 & 30.6 & 0.225 & 8.7 & 31.0 & 0.179 & 9.1 & 31.5 & 0.156 \\
    Per-Source Recipe & 0.00 & 8.0 & 32.4 & 0.177 & 8.8 & 31.3 & 0.177 & 9.3 & 31.5 & 0.165 \\
    Per-Cluster Recipe & $<$0.01 & 9.1 & 32.0 & 0.186 & -- & -- & -- & -- & -- & -- \\
    Image Embedding & $<$0.01 & \textbf{6.4} & 34.5 & 0.120 & 7.1 & 33.4 & 0.120 & 7.5 & 33.7 & 0.110 \\
    DAP~\cite{ahn2024impasto} & 0.17 & 7.5 & 33.5 & 0.137 & 7.6 & 32.9 & 0.128 & 8.2 & 33.1 & 0.112 \\
    LPIPS Cap & 0.67--2.67 & 9.0 & 32.0 & 0.135 & 7.0 & 34.2 & \textbf{0.090} & 8.9 & 32.7 & 0.113 \\
    \textbf{\ours{}} & 0.28 & \textbf{6.4} & \textbf{34.9} & \textbf{0.110} & \textbf{6.3} & \textbf{34.7} & \textbf{0.090} & \textbf{6.5} & \textbf{34.9} & \textbf{0.087} \\
    Direct Personalization & 71.7 & 5.2 & 36.7 & 0.077 & 5.5 & 36.3 & 0.075 & 5.8 & 36.1 & 0.083 \\
    \bottomrule
  \end{tabular*}
\end{table*}

\begin{figure*}[t]
  \centering
  \includegraphics[width=\linewidth]{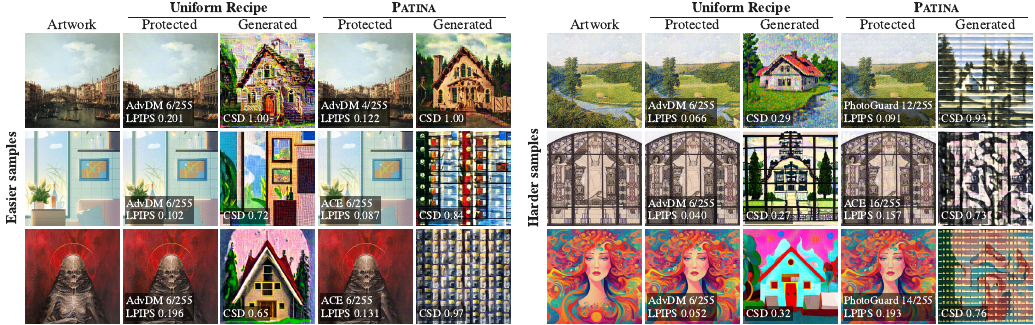}
  \par\vspace{-10pt}
  \caption{\textbf{Recipes for artworks from easy to hard.} Both strategies are set to a mean LPIPS of 0.15 under CSD, where uniform recipe gives these artworks AdvDM 6/255, and all generated images use the prompt ``an image of a house in $S^*$ style'' with the style token $S^*$. Labels give the recipe and LPIPS of each protected image and the CSD protection score of the artwork under that recipe, averaged over three seeds.}
  \label{fig:qual-compare}
  \vspace{-8pt}
\end{figure*}

\smallskip\noindent\textbf{Beyond the training setup.}
In practice, the predictor also needs to handle artworks and protection methods that are absent from its training setup.
\Cref{tab:general} leaves one source of artworks or one protection method out of training at a time.
When a new protection method appears, a protection tool can personalize a few artworks with it and use them to calibrate the overall level of every prediction (here, we use five artworks).
The \ours{} predictor remains second only to direct personalization in every column, so its advantage holds for sources and methods outside its training setup.

The imitator may also personalize in other ways, such as DreamBooth~\cite{ruiz2023dreambooth} or LoRA on another model.
Since the level of protection differs between these settings, each prediction, including mean report, is calibrated on artworks personalized in the new setting, leaving out the scored artworks.
After this calibration, the \ours{} predictor remains the most accurate in all three settings of \cref{tab:transfer}.
An artist also posts several artworks, so the imitator may personalize on several artworks (few-shot personalization).
For such a portfolio, our prediction is again the most accurate (\cref{tab:sets}).

\begin{takeaway}
\qbadge{2}~\textbf{Takeaway.} With little computation and no personalization, protection can be checked for each artwork before posting, nearly as accurately as one personalization run.
\end{takeaway}

\section{Choosing Protection per Artwork}
\label{sec:recipe}

Choosing a recipe for an artwork requires the predicted protection of every recipe, but the predictor of \cref{sec:predict} reads signals from an image already protected with the recipe in question.
Predicting every recipe would therefore require protecting the artwork with each recipe, which takes far more computation than one personalization.
Fortunately, the difficulty of an artwork is largely shared across recipes (\cref{sec:variation}), so one short trial protection is enough to predict the protection under every recipe.
We call the resulting method \ours{}.

\subsection{\ours{}}
\label{sec:method}

The protection scores of an artwork rise and fall together across recipes, so a cheap measurement that predicts some of them also predicts the rest, and \ours{} uses a short trial protection as this measurement.
\ours{} measures the image signals of the artwork and runs a trial protection with AdvDM~\cite{liang2023advdm} at 8/255 for 50 steps, recording its protection trace and protected image signals.
This setting is fixed before evaluation.
For each protection method, the \ours{} predictor maps these signals to predicted scores $\hat{s}(x,r)$ for the nine strengths.
\ours{} then gives the artwork the weakest recipe whose predicted score reaches a selection level $\gamma$,
\begin{equation}
  r^\star(x)=\operatorname*{arg\,min}_{r:\ \hat{s}(x,r)\ge\gamma}\ \varepsilon(r),
  \label{eq:choose}
\end{equation}
and among recipes of equal strength it takes the method with the highest prediction.
The weakest such recipe keeps the perturbation as invisible as the prediction allows, and raising $\gamma$ trades visibility for a larger margin of protection.
An artist can also bound the visibility of every artwork, and \ours{} then chooses only among recipes whose predicted LPIPS stays within the limit and steps down the strength when the measured LPIPS exceeds it.

If no recipe, up to the strongest one within the visibility limit, is predicted to pass the threshold, \ours{} warns the artist in advance.
Every such recipe is then likely to leave the artwork below the threshold, so the artist can decide whether to post it at all.
The warning is especially useful when the imitator purifies the protected image, since strong purification leaves many artworks below the threshold under every recipe (\cref{tab:purify-app}).
An artist who expects a purification, such as JPEG compression, can turn on the purification-aware mode, in which the same rule runs on a predictor trained on the protection scores after that purification.

\subsection{Results}

\label{sec:results}
\begin{table}[t]
  \centering
  \caption{\textbf{Protection under purification.} The imitator purifies the protected image before personalization, and at the 50\% threshold both strategies choose recipes for the score after purification (purification-aware). In the middle block, every artwork gets the strongest strength 18/255, with the best single method in the first row and the method that \ours{} predicts to protect best in the other two. These two rows split the artworks by whether \ours{} warns that no recipe is predicted to pass ($\max_r\hat{s}<\tau$), and the bottom block gives the warned percentage under each purification.}
  \label{tab:purify-app}
  \footnotesize
  \setlength{\tabcolsep}{0pt}
  \begin{tabular*}{\linewidth}{@{\extracolsep{\fill}}lccc@{}}
    \toprule
    Pass Rate (\%) & CSD\,$\uparrow$ & \artsim{}\,$\uparrow$ & CLIP-IQA\,$\uparrow$ \\
    \midrule
    \multicolumn{4}{@{}l}{\emph{JPEG, purification-aware, mean LPIPS limit 0.20}} \\
    Uniform Recipe & 79.0 & 67.4 & 62.5 \\
    \textbf{\ours{}} & \textbf{82.4} & \textbf{74.4} & \textbf{66.7} \\
    \midrule
    \multicolumn{4}{@{}l}{\emph{Noisy Upscaling~\cite{honig2025adversarial}, 18/255}} \\
    One Recipe for All Artworks & 54.0 & 47.0 & 55.0 \\
    \textbf{\ours{}}, Not Warned Artworks & 73.8 & 69.4 & 63.2 \\
    \textbf{\ours{}}, Warned Artworks & 33.3 & 31.2 & 41.3 \\
    \midrule
    \multicolumn{4}{@{}l}{\emph{Warned Artworks (\%)}} \\
    JPEG & 2.8 & 2.6 & 0.0 \\
    Noisy Upscaling & 48.0 & 56.8 & 40.0 \\
    \bottomrule
  \end{tabular*}
\end{table}

\smallskip\noindent\textbf{Protocol.}
We compare \ours{} with uniform recipe, which gives every artwork the weakest recipe whose pass rate on the training artworks reaches a target, as current protection tools do, and with strategies that adapt the recipe to each artwork.
Image embedding, DAP, and direct personalization (\cref{sec:signals}) choose with \cref{eq:choose} from their predicted or measured scores, and LPIPS cap protects every artwork with one method at the strongest strength within a common LPIPS limit.
Direct personalization personalizes each recipe once and shows what this costly measurement achieves.
Each strategy has one knob that traces a curve of pass rate against mean LPIPS (\cref{fig:q3,fig:curves-all}), and \cref{tab:main} shows three views of them.

\smallskip\noindent\textbf{Strategy comparison.}
At the 50\% threshold, every strategy that measures each artwork improves on uniform recipe in all three views and on all three scores (upper block of \cref{tab:main}), while giving one recipe to each source or to each style cluster does not improve consistently on it (\cref{tab:visibility}).
The recipe average that protection tools follow therefore leaves a large gain unused.
Among the strategies that need no personalization, \ours{} recovers the most (\cref{tab:main}), at about 1/250 of the cost of direct personalization.
It reaches the same pass rate as uniform recipe with about half the visibility, and the $L_\infty$ strength and the PSNR show the same trend (\cref{tab:visibility}).
LPIPS cap protects each artwork several times and never predicts the protection score, so it tells the artist nothing about protection.
The ranking holds at the 80\% threshold (lower block of \cref{tab:main}), and \ours{} stays ahead of uniform recipe at every threshold (\cref{fig:sweep}) and when the recipe chosen for one score is evaluated with another of the three protection scores (\cref{tab:cross}).

\begin{table}[t]
  \centering
  \caption{\textbf{Protection against few-shot personalization.} The imitator personalizes one model on a portfolio of four or five artworks of one artist. The first column group gives the RMSE of predicting the protection score of each portfolio from the mean prediction over its artworks, where uniform recipe uses mean report. LPIPS cap chooses recipes by LPIPS alone without predicting the protection score, so it has no RMSE (dash). The second column group gives the percentage of portfolios below the 50\% threshold ($s<\tau$) when each strategy chooses recipes for their artworks at a mean LPIPS limit of 0.15. Bold marks the best result in each column.}
  \label{tab:sets}
  \footnotesize
  \setlength{\tabcolsep}{0pt}
  \begin{tabular*}{\linewidth}{@{\extracolsep{\fill}}lcccc@{}}
    \toprule
    & \multicolumn{2}{c}{RMSE} & \multicolumn{2}{c}{Portfolios with $s<\tau$ (\%)} \\
    \cmidrule(lr){2-3}\cmidrule(lr){4-5}
    Strategy & CSD\,$\downarrow$ & \artsim{}\,$\downarrow$ & CSD\,$\downarrow$ & \artsim{}\,$\downarrow$ \\
    \midrule
    Uniform Recipe & 0.246 & 0.231 & 25.4 & 16.4 \\
    Image Embedding & 0.230 & 0.225 & 20.0 & 22.4 \\
    DAP~\cite{ahn2024impasto} & 0.226 & 0.225 & 20.0 & 8.1 \\
    LPIPS Cap & -- & -- & 19.3 & 24.4 \\
    \textbf{\ours{}} & \textbf{0.180} & \textbf{0.217} & \textbf{14.8} & \textbf{5.8} \\
    \bottomrule
  \end{tabular*}
\end{table}

\ours{} gains protection by giving less strength to the artworks that need little and more to the artworks that need much (\cref{fig:q3}c,d and \cref{fig:qual-compare}).
Uniform recipe leaves the hardest group far below the threshold and gives the easiest group the most visible perturbation, while \ours{} makes the perturbation of the easiest groups less visible without lowering their pass rate and brings the pass rate of the hardest group close to those of the other groups.
Among the hardest quarter of the artworks, uniform recipe leaves almost half below the threshold under the two style scores, and \ours{} leaves few (\cref{tab:main}).
An artist may also bound the visibility of every artwork, and under a per-artwork LPIPS limit of 0.25, \ours{} again leaves the fewest artworks below the threshold on all three protection scores (\cref{tab:perartwork}).

\smallskip\noindent\textbf{Purification and early warning.}
An imitator who obtains the protected image may purify it before malicious personalization~\cite{nie2022diffpure,cao2023impress,zhao2024gridpure,honig2025adversarial}, and we test the common JPEG compression and the stronger Noisy Upscaling~\cite{honig2025adversarial}.
JPEG compression removes weak perturbations first, which lowers the protection of easy artworks under recipes chosen for clean personalization.
The purification-aware mode of \cref{sec:method} keeps the rule of \cref{eq:choose} and only replaces the predictor with one trained on the protection scores after purification, so the same tool adapts to a known purification.
With this mode, \ours{} lets more artworks pass the threshold than uniform recipe chosen for the same score after JPEG, at every limit that both reach (top block of \cref{tab:purify-app} and \cref{tab:jpeg}).

Noisy Upscaling removes most of the protection, and even the best recipe for all artworks at 18/255 lets only about half of the artworks pass the threshold, so no choice of recipe lets every artwork pass the threshold.
The artist then needs to know which artworks stay below the threshold under every recipe, and \ours{} warns about an artwork whenever no recipe is predicted to pass the threshold after purification.
Under CSD, only 33\% of the warned artworks pass the threshold even at 18/255, against 74\% of the others (middle block of \cref{tab:purify-app}), while under JPEG compression \ours{} warns about fewer than 3\% of the artworks (bottom block).
The warning therefore tells the artist before posting which artworks are likely to stay below the threshold, and a protection method that resists purification can reduce the number of these artworks once the predictor learns its scores.

\smallskip\noindent\textbf{Few-shot personalization.}
An imitator often personalizes one model on an artist's portfolio.
Even in this scenario, \ours{} needs no retraining, since it predicts each artwork of the portfolio from its own trial protection and gives each artwork its own recipe.
The mean of these predictions predicts the protection score of the portfolio more accurately than mean report, DAP, and image embedding (\cref{tab:sets}).
With these recipes, \ours{} also leaves the fewest portfolios below the threshold on both style scores.
Choosing per artwork therefore extends from single artworks to the portfolios that an imitator collects from one artist.

\subsection{Analysis}
\label{sec:analysis}

The last row of \cref{tab:ablation} tests the principle of \ours{}, that one trial protection can replace protecting the artwork with every recipe.
Protecting the artwork with all 54 recipes and predicting each recipe from its own protected image chooses nearly as well at about 80 times the cost, so the shared structure of \cref{sec:variation} lets one trial stand for all recipes.
Among the signals, those of the artwork alone already choose recipes far better than uniform recipe, and under CSD and \artsim{} the protected image signals of the trial add most of the remaining gain, the same division of information as in the prediction of \cref{sec:predict}.
The length of the trial hardly matters, and a 10-step trial lowers the artist-side cost to 15\% of one personalization.
Any recipe also works as the trial, and most of the gain remains on sources held out from training (\cref{app:trial,app:unseen-choice}).

Choosing the method for each artwork matters most under CSD, where giving every artwork the same method, the best one on the training artworks, lowers the pass rate from 96.2\% to 88.6\% (\cref{tab:ablation}).
Under CSD, \ours{} spreads its choices over AdvDM, PhotoGuard, and ACE, since the best method differs from one artwork to another.
Under \artsim{} and CLIP-IQA, it gives one method to most artworks, so fixing the method costs little there.
The recipe set can grow, since a new method raises the pass rate once the \ours{} predictor also learns its scores (\cref{app:unseen-choice}).

\begin{takeaway}
\qbadge{3}~\textbf{Takeaway.} With existing protection methods, more artworks pass the protection threshold with less visible perturbations once the recipe is chosen for each artwork, and one short trial protection is enough to make this choice before posting.
\end{takeaway}

\begin{table}[t]
  \centering
  \caption{\textbf{Analysis of \ours{}.} We change one part of \ours{} at a time under the 50\% threshold. The first block starts from uniform recipe, which uses no signal of the artwork, and adds one signal category at a time in the order it becomes available, ending at \ours{}. The second block changes the number of optimization steps of the trial protection. Same method gives every artwork the method with the highest pass rate on the training artworks, and \ours{} chooses only the strength. All 54 recipes protected takes the signals of each recipe from the image protected with it, at about 80 times the cost of \ours{}.}
  \label{tab:ablation}
  \footnotesize
  \setlength{\tabcolsep}{0pt}
  \begin{tabular*}{\linewidth}{@{\extracolsep{\fill}}lccc@{}}
    \toprule
    Pass Rate (\%), Mean LPIPS Limit 0.15 & CSD\,$\uparrow$ & \artsim{}\,$\uparrow$ & CLIP-IQA\,$\uparrow$ \\
    \midrule
    Uniform Recipe & 82.2 & 81.6 & 89.9 \\
    Image Signals & 92.8 & 97.1 & 95.6 \\
    + Trial Protection Trace & 92.4 & 97.3 & 96.3 \\
    + Trial Protected Image (\textbf{\ours{}}) & 96.2 & 98.3 & 95.7 \\
    \midrule
    Trial Length (Steps) = 10 & 95.0 & 98.2 & 96.1 \\
    Trial Length (Steps) = 25 & 95.3 & 98.3 & 96.5 \\
    Trial Length (Steps) = 50 (\textbf{\ours{}}) & 96.2 & 98.3 & 95.7 \\
    Trial Length (Steps) = 100 & 94.7 & 98.3 & 95.8 \\
    \midrule
    \textbf{\ours{}}, Same Method & 88.6 & 98.0 & 95.4 \\
    \textbf{\ours{}}, All 54 Recipes Protected & 94.9 & 98.5 & 97.0 \\
    \bottomrule
  \end{tabular*}
  \vspace{-1.3pt}
\end{table}

\section{Discussion and Conclusion}
\label{sec:discussion}

\smallskip\noindent\textbf{Implications for evaluation.}
Evaluating protection by the average of a recipe hides the artworks that the recipe leaves below the protection threshold.
Averaged over recipes, artworks differ in protection 1.6 to 1.9 times as much as personalization seeds do, and the difficulty of an artwork is largely shared across protection methods (\cref{sec:variation,app:seed}).
A ranking of methods by their averages still does not transfer to every artwork, since giving every artwork the method that is best on average lowers the pass rate under CSD (\cref{tab:ablation}).
Seed noise also remains, so a single personalization run is an uncertain check of one artwork.
Evaluations of protection methods should report, next to the recipe average, the distribution over artworks and the noise across seeds.

\smallskip\noindent\textbf{Implications for protection tools.}
Protection becomes a decision made for each artwork.
A tool can tell the artist, at a fraction of the cost of one personalization, how well the artwork will be protected and which recipe to apply.
It gives most artworks a weaker perturbation than a single recipe would, and it gives the hard artworks the strength they need.
When a purification removes most of the protection, the tool still tells the artist before posting for which artworks no recipe is predicted to pass the threshold, so the artist can decide how to share each of them.
The protection threshold is also a choice of the artist.
We evaluate with the 50\% and 80\% thresholds that five raters set, and each artist may set a tighter or looser threshold for their own artworks.
Since \ours{} predicts the protection score itself, the same predictor serves any threshold, and both the recipe and the warning follow the threshold that the artist sets.
The structure shared across recipes explains why one trial protection is enough to adapt both the protection method and the strength.

\smallskip\noindent\textbf{Open questions.}
Two questions remain open, and image protection research is only beginning to address them.
First, which protection methods resist strong purification such as Noisy Upscaling~\cite{honig2025adversarial}?
Adding them to the recipe set would reduce the number of artworks that \ours{} warns about.
Second, does the structure shared across recipes hold for protection methods designed for newer architectures such as Flux~\cite{blackforestlabs2025kontext} and Qwen-Image~\cite{wu2025qwenimage}, which replace the U-Net with a diffusion transformer?
The predictor extends to methods that are absent from training (\cref{sec:predict-results}), and a new method raises the pass rate of \ours{} once the predictor learns its scores (\cref{sec:analysis}), which suggests that such methods can join the recipe set of \ours{}.

\smallskip\noindent\textbf{Conclusion.}
The same recipe protects artworks to very different degrees, and this variation is a property of the artwork that persists across seeds, visibility, personalization settings, and portfolios.
Cheap signals predict it before personalization, nearly as accurately as one personalization run.
Because the difficulty of an artwork is largely shared across recipes, one short trial protection is enough to choose a recipe for each artwork, and more artworks pass the protection threshold at the same visibility than under any single recipe or other per-artwork strategy that skips personalization.
An artist can therefore protect each artwork just enough and learn in advance when no recipe is predicted to pass.

\clearpage
\section*{LLM Usage Statement}
We used Claude Code with Claude Opus 5.5 and 5.1 in this work.
Claude Code wrote the code and ran and orchestrated the experiments, and the authors reviewed and approved the experiments and the analyses of their results.
The authors wrote the initial draft of the manuscript, and Claude Code assisted with the literature survey and with polishing the text and the figures.
We also obtained reviewer-style feedback on drafts of the manuscript from Claude and ChatGPT.
The research ideas, claims, and conclusions are those of the authors.
The authors take full responsibility for all submitted content, regardless of any LLM or AI-tool use.

{
    \small
    \bibliographystyle{ieeenat_fullname}
    \bibliography{main}
}

\clearpage
\appendix
\renewcommand{\thefigure}{A\arabic{figure}}\renewcommand{\thetable}{A\arabic{table}}
\flushbottom\setlength{\parskip}{0pt plus 1pt}
\setcounter{totalnumber}{5}
\renewcommand{\topfraction}{0.9}
\renewcommand{\dbltopfraction}{0.9}

\section{Related Work}
\label{app:related}

\subsection{Style Personalization}

Personalization adapts a text-to-image diffusion model to a new concept or style from images that a user provides.
Few-shot methods learn from several images of the concept.
DreamBooth~\cite{ruiz2023dreambooth} fine-tunes the denoiser, Textual Inversion~\cite{gal2023textual} learns a new word embedding from a few images, and Custom Diffusion~\cite{kumari2023multi} tunes a few parameters of the text conditioning.
Single-image methods learn a style from one reference image.
StyleDrop~\cite{sohn2023styledrop} tunes a small part of the base model and refines it with feedback, DreamStyler~\cite{ahn2024dreamstyler} optimizes a multi-stage textual embedding, B-LoRA~\cite{frenkel2024blora} separates style and content within the LoRA~\cite{hu2022lora} weights of two SDXL~\cite{podell2024sdxl} blocks, and ZipLoRA~\cite{shah2024ziplora} merges a style LoRA trained on one exemplar with a subject LoRA.
Our imitator follows this setting with LoRA~\cite{hu2022lora} on SD~2.1, so each outcome of protection belongs to a single artwork.

\subsection{Protection against Style Mimicry}

\smallskip\noindent\textbf{Protection methods.}
Perturbation-based protections against the personalization of diffusion models raise the denoising loss of the U-Net~\cite{liang2023advdm,liang2023mist,vanle2023antidreambooth,liu2024metacloak,styleprotect2026}, pull its predicted noise toward a target~\cite{zheng2025ace}, shift the latent of the image encoder~\cite{salman2023photoguard,li2024pid}, or reverse the gradient of score distillation~\cite{xue2024sds}.
Other protections mislead the cross-attention layers that fine-tuning updates~\cite{xu2024caat}, amplify the error of latent encoding and weaken the guidance~\cite{qiu2026void}, or target style features~\cite{shan2023glaze,li2025styleguard}.
Data poisoning~\cite{shan2024nightshade} corrupts a concept when a model trains on many scraped images, a threat different from personalization on the artworks of one artist.

\smallskip\noindent\textbf{Newer generators.}
The diffusion transformer~\cite{peebles2023scalable} replaces the U-Net in many recent text-to-image generators~\cite{esser2024scaling,blackforestlabs2025kontext,wu2025qwenimage}, and autoregressive generators predict image tokens one by one~\cite{sun2024llamagen} or scale by scale~\cite{tian2024var}.
Protection for these generators is at an early stage, and only a handful of studies address it.
These studies disrupt the attention through which in-context editors such as Flux.1 Kontext read the reference image~\cite{shen2025decontext,grebe2026veto}.
A protection is also hard to transfer between U-Net models and these generators, because the two differ both in architecture and in the training signal.
The newer generators mix image and text tokens in joint attention blocks, and many of them learn with flow matching~\cite{lipman2023flow,esser2024scaling}, while the U-Net models in our study learn to predict the added noise.
Newer editors thus often circumvent protections designed for earlier diffusion pipelines~\cite{grebe2026veto}.
We therefore build the recipe set on U-Net models, for which protection research is active and many protection methods are available.
Whether our findings extend to these generators remains an open question that we raise in the discussion (\cref{sec:discussion}).

\smallskip\noindent\textbf{Adaptive protection.}
Some protections adapt the perturbation to the input at a fixed strength.
Glaze~\cite{shan2023glaze} picks a target style for each artist, Impasto~\cite{ahn2024impasto} distributes the perturbation over the image with a perceptual model and suggests a difficulty estimate that we use as the DAP baseline, and FastProtect~\cite{ahn2025fastprotect} selects one of several pretrained perturbations for each image.
Hönig \etal~\cite{honig2025adversarial} show that simple purification removes existing protections and that the protection is highly non-uniform across artists.
Our work measures the outcome of protection for each artwork, predicts it in advance, and uses the prediction to choose both the method and the strength of the protection.

\smallskip\noindent\textbf{Purification.}
An imitator can remove a protective perturbation before personalization.
Conventional purification applies a common image transformation, and JPEG compression is the one that many protection studies test.
Diffusion-based purification uses a diffusion model or its autoencoder to remove the perturbation.
DiffPure~\cite{nie2022diffpure} adds a small amount of noise and denoises the image with a diffusion model, IMPRESS~\cite{cao2023impress} optimizes the image to agree with its reconstruction by the latent autoencoder, and GrIDPure~\cite{zhao2024gridpure} applies diffusion purification to small grids of the image to keep its structure.
Hönig \etal~\cite{honig2025adversarial} propose Noisy Upscaling, a very simple but effective purification that adds Gaussian noise and enlarges the image with the SD upscaler.
Off-the-shelf image-to-image models such as Flux.1~\cite{blackforestlabs2025kontext} also remove several protections with a short prompt~\cite{pleimling2026offtheshelf}.

These diffusion-based purifications remove most of the current protections~\cite{honig2025adversarial,pleimling2026offtheshelf}.
The goal of many protections is to raise the effort of a malicious imitator, so that mimicry succeeds only after the imitator finds and applies such a purification.
We test protection under a weak purification, JPEG compression, and under a strong one, Noisy Upscaling.
Under the strong one, \ours{} warns the artist in advance about the artworks for which no recipe is predicted to pass the protection threshold (\cref{sec:recipe}).

\section{Experimental Setup}
\label{app:setup}

\begin{figure*}[t]
  \centering
  \includegraphics[width=\linewidth]{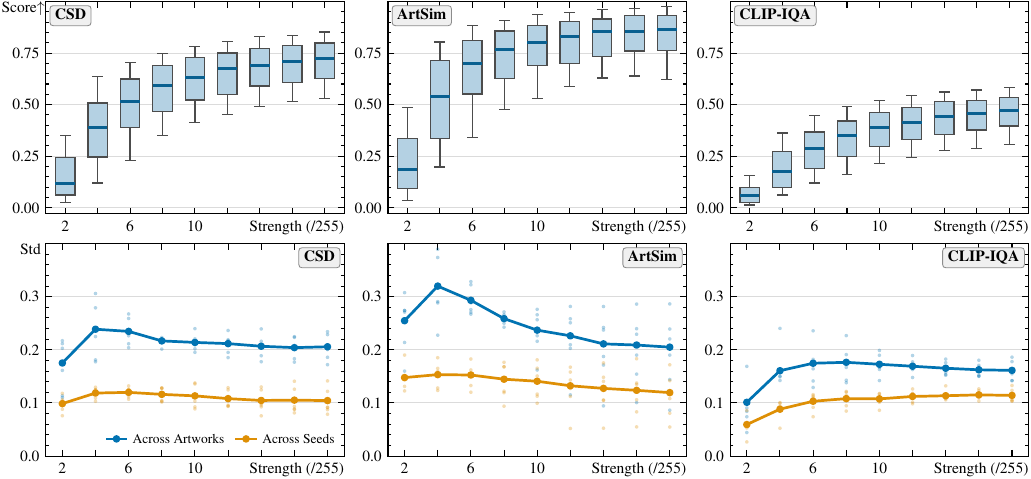}
  \vspace{-20pt}
  \caption{\textbf{Spread of protection for the three scores.} The top row gives the protection score of 300 artworks at each strength, and the bottom row the standard deviation within one recipe across artworks and across seeds on 100 artworks, as in \cref{fig:variation}a and~b.}
  \label{fig:spread-all}
\end{figure*}

\subsection{Protection Recipes}

The recipes combine six protection methods, AdvDM~\cite{liang2023advdm}, PhotoGuard~\cite{salman2023photoguard}, ACE~\cite{zheng2025ace}, SDS($-$)~\cite{xue2024sds}, StyleProtect~\cite{styleprotect2026}, and StyleGuard~\cite{li2025styleguard}.
The nine strengths run from 2/255 to 18/255 in steps of 2/255.
Each method optimizes the perturbation on SD~2.1~\cite{rombach2022ldm} with projected gradient descent for 100 steps of size 1/255 under $\|\delta\|_\infty\le\varepsilon$, with pixel values in $[0,1]$.
The visibility of the perturbation is $\mathrm{LPIPS}(x,\tilde{x})$~\cite{zhang2018lpips}, the price the artist pays for protection.

\subsection{Data, Personalization, and Purification}

\smallskip\noindent\textbf{Data.}
The 300 artworks are 200 paintings from WikiArt~\cite{saleh2015wikiart}, 50 images from StyleBench~\cite{gao2024styleshot}, and 50 style reference images from OmniStyle~\cite{wang2025omnistyle}, all resized to $512\times512$.
Every artwork is protected with all 54 recipes, and each protected image is personalized.
The 100 validation artworks are 66 from WikiArt, 17 from StyleBench, and 17 from OmniStyle.

\smallskip\noindent\textbf{Personalization.}
The imitator personalizes SD~2.1 on one artwork with LoRA~\cite{hu2022lora} at rank 4, learning rate $10^{-4}$, and 300 steps.
The 20 generated images are four images for each of five prompts, ``an image of a house in $S^*$ style'', ``an image of a tree in $S^*$ style'', ``an image of a woman in $S^*$ style'', ``an image of a city in $S^*$ style'', and ``an image of a boat in $S^*$ style'', where $S^*$ is the identifier token of the style.
The base model generates images from the same prompts with the same generation seed.
Prior protection work evaluates against the same kind of personalization, DreamBooth or LoRA on SD~\cite{liang2023mist,zheng2025ace,li2025styleguard,styleprotect2026,ahn2024impasto,ahn2025fastprotect}.
Protection methods are designed and studied mostly on this model family~\cite{liang2023advdm,vanle2023antidreambooth,liu2024metacloak,xu2024caat,qiu2026void}, and every method of our recipes is built on it.

\smallskip\noindent\textbf{Other personalization settings.}
SD~1.5 and SDXL~\cite{podell2024sdxl} serve as further personalization models.
For them, we protect the 100 validation artworks with AdvDM, PhotoGuard, and StyleGuard at 16/255.
With the same artworks and protected images, we also fine-tune the whole denoiser of SD~2.1 with DreamBooth~\cite{ruiz2023dreambooth} at learning rate $5\times10^{-6}$ and 500 steps.

\smallskip\noindent\textbf{Few-shot personalization.}
In few-shot personalization, the imitator trains one model on a portfolio of several artworks by one artist.
We take 20 portfolios from the data of StyleProtect~\cite{styleprotect2026}, 15 WikiArt artists with five artworks each and five further portfolios of four or five artworks, 98 artworks in total.
Each portfolio is personalized together on SD~2.1 with LoRA at rank 4, learning rate $10^{-4}$, and 500 steps.
The portfolios are protected under ACE and AdvDM, the two methods with the highest pass rate under CSD in the main experiment, at the nine strengths, which gives 18 recipes.
The style similarity of a generated image is measured against the mean embedding of the artworks in the portfolio.

\smallskip\noindent\textbf{Purification.}
The imitator may apply a purification $T$ before personalization and train on $T(\tilde{x})$.
JPEG compression at quality 75 is a conventional purification that protection work commonly tests.
Noisy Upscaling~\cite{honig2025adversarial} adds Gaussian noise and enlarges the image with the SD upscaler, and it is the most effective of the purifications that Hönig \etal compare.
The main experiments use no purification.
The purification runs use the 100 validation artworks, AdvDM, PhotoGuard, and StyleGuard at seven strengths from 6/255 to 18/255.

\subsection{Evaluation}
\label{app:score}

\begin{figure*}[t]
  \centering
  \includegraphics[width=\linewidth]{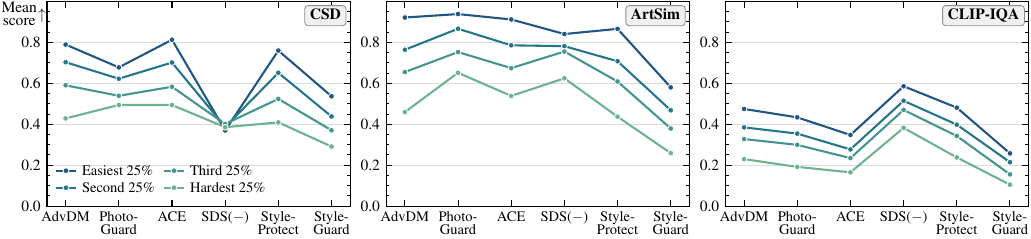}
  \vspace{-20pt}
  \caption{\textbf{Groups of artworks keep their order across methods.} For each method, the artworks are split into four groups by their mean score under the other five methods, and each point is the mean score of a group under the method over the nine strengths and three seeds.}
  \label{fig:order-all}
\end{figure*}

\smallskip\noindent\textbf{Protection scores.}
Each protection score rescales a metric of the generated images to the range from 0 to 1, where 0 means that the protected personalization imitates the artwork as well as the original one and 1 means that it imitates the artwork no better than the base model, or, for the CLIP-IQA score, that its images have zero quality.
The raw similarity of an artwork depends on how close the base model already comes to its style and on how well the original personalization learns it, and the rescaling puts every artwork on one scale, so that one threshold applies to all artworks.
For an artwork $x$ and a recipe $r$, the original personalization, the protected personalization, and the base model each generate 20 images from the same prompts.
The style similarity of generated images $g_1,\dots,g_{20}$ to the artwork is
\begin{equation}
  S=\frac{1}{20}\sum_{i=1}^{20}\cos\big(\phi(g_i),\phi(x)\big)
\end{equation}
for a feature extractor $\phi$.
We write $S_o$, $S_p$, and $S_b$ for the similarity of the images from the original personalization, the protected personalization, and the base model.
The protection score is
\begin{equation}
  s(x,r)=\mathrm{clip}_{[0,1]}\!\left(\frac{S_o-S_p}{S_o-S_b}\right),
\end{equation}
which is 0 when $S_p\ge S_o$ and 1 when $S_p\le S_b$.

The CSD score uses CSD~\cite{somepalli2024csd}, a CLIP ViT-L trained with contrastive learning on images labeled with style tags.
The \artsim{} score uses the Inception network of ArtFID~\cite{wright2022artfid}, which is trained to classify the artist and the stylistic period of about 250K artworks, and takes its features before the classification heads.
The CLIP-IQA score uses the CLIP-IQA quality $Q$ of the generated images~\cite{wang2023clipiqa}, with $s=\mathrm{clip}_{[0,1]}\big((Q_o-Q_p)/Q_o\big)$ in the same way.

\smallskip\noindent\textbf{Thresholds.}
How much protection is enough differs from artist to artist, and the protection threshold $\tau$ sets one reference level for the experiments of this paper.
An artwork passes the threshold when $s\ge\tau$, and each threshold marks the score at which a given percentage of the protected generations no longer show the style of the artwork or are too distorted to use.
In our study, five raters set two thresholds for each score, the 50\% threshold, where about 50\% of the protected generations do so, and the 80\% threshold, where about 80\% of them do.
We use these two thresholds for quantitative evaluation.
However, each artist can freely set a tighter or looser threshold, since the \ours{} predictor predicts the protection score itself.

\begin{figure*}[t]
  \centering
  \includegraphics[width=\linewidth]{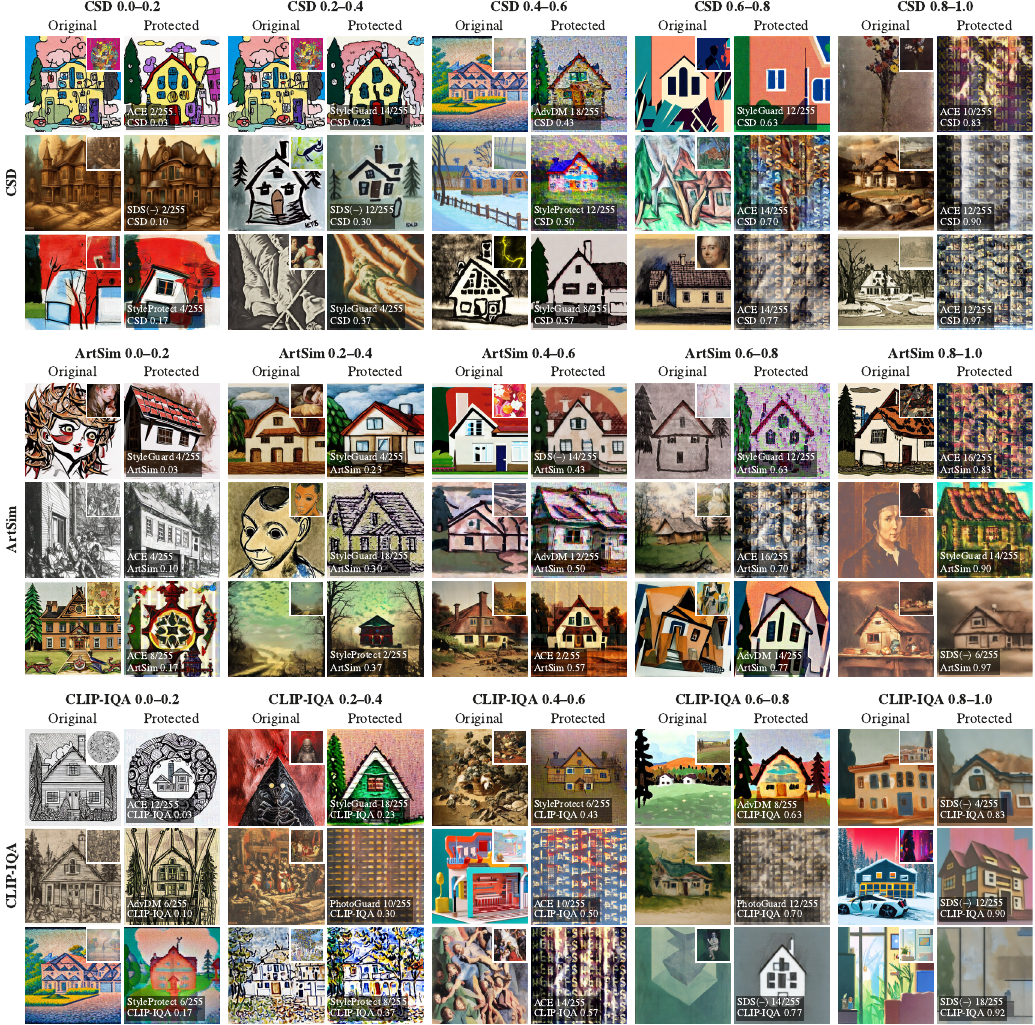}
  \caption{\textbf{Cases in each range of the protection scores.} Each case shows an image generated after personalization on the artwork (Original, with the artwork in its corner) and an image generated with the same prompt after personalization on the protected image (Protected, with the recipe and the protection score). All generated images use the prompt ``an image of a house in $S^*$ style''.}
  \label{fig:qual-bins}
\end{figure*}

For each score, we split the protection score into 20 ranges of width 0.05 and draw 30 random cases per range, each with the artwork, an image from the original personalization, and the image from the protected personalization with the same prompt.
The raters, four of them for CLIP-IQA, view these cases with the method and strength hidden and choose two ranges for each score, where about 50\% and about 80\% of the protected generations lose the style or become too distorted to use.
Each threshold is the median of the lower bounds of the chosen ranges, and with an even number of raters it is the higher of the two middle values.
At the 50\% threshold, these lower bounds range over 0.35--0.60 for CSD, 0.45--0.60 for \artsim{}, and 0.10--0.30 for CLIP-IQA.
The two thresholds are therefore 0.40 and 0.65 for CSD, 0.50 and 0.70 for \artsim{}, and 0.20 and 0.40 for CLIP-IQA.
\Cref{fig:qual-bins} shows what these thresholds look like, with three cases in each range of width 0.2 for every score, namely the cases whose scores lie closest to three evenly spaced points.

\smallskip\noindent\textbf{Cross-validation.}
Methods that learn from data are trained with 5-fold cross-validation over artworks, repeated five times with different splits.
They are scored on the 100 validation artworks, each personalized in three runs, and each artwork is scored by the fold that holds it out.
A pass rate averages, for each artwork, the fraction of the three runs that pass the threshold, and an RMSE is computed in each run and averaged over the runs.
Each reported value is the mean over the five repetitions of the split.

\section{Unequal Protection across Artworks}
\label{app:variation}

This section gives the details and further results of \cref{sec:variation}, which answers Q1 of the introduction, and asks whether seeds, visibility, personalization settings, or the kind of artwork explain the spread across artworks.

\begin{table}[t]
  \centering
  \caption{\textbf{Matching visibility leaves the difference between artworks.} We compare the spread of protection at one strength with the spread at one LPIPS on 100 artworks. The first four columns give the standard deviation of the protection score across artworks and across seeds, averaged over the six methods as the square root of the mean variance. Fixed strength uses the strength with mean LPIPS closest to 0.10, and matched LPIPS interpolates each artwork to LPIPS 0.10. The last column gives the percentage of similar-texture artwork pairs whose seed-averaged scores differ by over 0.3, while seed noise alone gives at most 0.8\%.}
  \label{tab:matched}
  \footnotesize
  \setlength{\tabcolsep}{0pt}
  \begin{tabular*}{\linewidth}{@{\extracolsep{\fill}}lccccc@{}}
    \toprule
    & \multicolumn{2}{c}{Fixed Strength} & \multicolumn{2}{c}{Matched LPIPS} & Similar \\
    \cmidrule{2-3}\cmidrule{4-5}
    Score & Artworks & Seeds & Artworks & Seeds & Pairs (\%) \\
    \midrule
    CSD & 0.230 & 0.121 & 0.195 & 0.114 & 18.7 \\
    \artsim{} & 0.287 & 0.154 & 0.231 & 0.138 & 25.9 \\
    CLIP-IQA & 0.174 & 0.104 & 0.156 & 0.097 & 10.0 \\
    \bottomrule
  \end{tabular*}
\end{table}

\begin{figure*}[t]
  \centering
  \includegraphics[width=\linewidth]{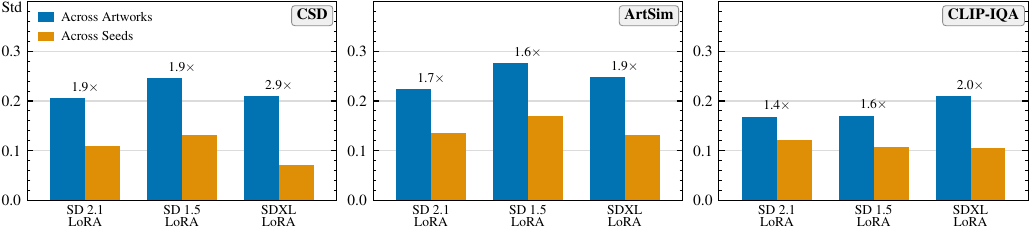}
  \vspace{-20pt}
  \caption{\textbf{Artworks differ more than seeds in other personalization models.} Standard deviation of the protection score within one recipe across artworks and across seeds, with their ratio above each pair, for AdvDM, PhotoGuard, and StyleGuard at 16/255.}
  \label{fig:backbones}
\end{figure*}

\subsection{Do Artworks Differ More than Seeds?}
\label{app:seed}

\Cref{fig:spread-all} extends \cref{fig:variation}a and~b to the three scores.
Under every score, the protection score at one strength spreads widely across artworks (top row).
The spread across artworks is 1.6 to 1.9 times the spread across seeds (bottom row), and it exceeds the spread across seeds in every one of the 162 combinations of six protection methods, nine strengths, and three scores.

\subsection{Does Visibility Explain the Spread?}
\label{app:visibility}

At the same $L_\infty$ strength, a perturbation is more visible on an artwork with simple texture, so one might suspect that the spread reflects visibility alone.
However, the spread remains when the visibility is matched.
Measuring the protection score of each artwork at a common LPIPS of 0.10, interpolated over strength, lowers the spread across artworks by only 10 to 19\%, and the spread stays 1.6 to 1.7 times the seed noise (\cref{tab:matched}).
Among artwork pairs with the same LPIPS and similar texture, 10 to 26\% differ by more than 0.3 in their seed-averaged scores, where seed noise alone produces at most 0.8\%.
For example, two artworks with similar LPIPS and texture under SDS($-$) have CSD protection scores as different as 0.95 and 0.02.
Visibility therefore explains only a small part of the difference between artworks, and \cref{app:hard} examines which other properties of the artwork matter.

\begin{figure}[t]
  \centering
  \includegraphics[width=\linewidth]{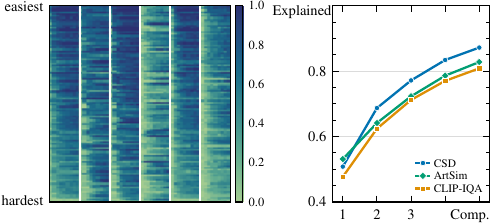}
  \vspace{-20pt}
  \caption{\textbf{Difficulty is shared across recipes.} Left, seed-averaged CSD protection scores (color) of 100 artworks, sorted from the easiest to the hardest, under the 54 recipes, grouped by method with strength increasing to the right. Right, the share of variation explained by the leading components after the mean of each recipe is removed from the matrix.}
  \label{fig:structure}
\end{figure}

\begin{table}[t]
  \centering
  \caption{\textbf{Spread of protection across portfolios.} Standard deviation of the protection score within a recipe across 20 portfolios and 300 single artworks under the same 18 recipes.}
  \label{tab:sets-spread}
  \footnotesize
  \setlength{\tabcolsep}{0pt}
  \begin{tabular*}{\linewidth}{@{\extracolsep{\fill}}lcc@{}}
    \toprule
    Standard Deviation & CSD & \artsim{} \\
    \midrule
    Across Portfolios & 0.239 & 0.223 \\
    Across Single Artworks & 0.220 & 0.244 \\
    \bottomrule
  \end{tabular*}
\end{table}

\subsection{Is Difficulty Shared across Recipes?}
\label{app:shared}

\Cref{fig:order-all} shows the groups of \cref{fig:variation}c for the three scores, and every method keeps the order of the groups under \artsim{} and CLIP-IQA.
Under CSD, SDS($-$) gives the weakest protection of the six methods, and its four groups lie close together within 0.37--0.40.
For the structure of the full matrix, we arrange the seed-averaged scores of 100 artworks under the 54 recipes into a matrix, remove the mean of each recipe, and apply a singular value decomposition (\cref{fig:structure}).
One component explains 48 to 53\% of the variation across the three scores, and three components explain 71 to 77\%.
The structure does not come from the order of strengths, since at a fixed strength one component explains on average 56 to 63\% of the variation across the six methods.

\begin{figure*}[t]
  \centering
  \includegraphics[width=\linewidth]{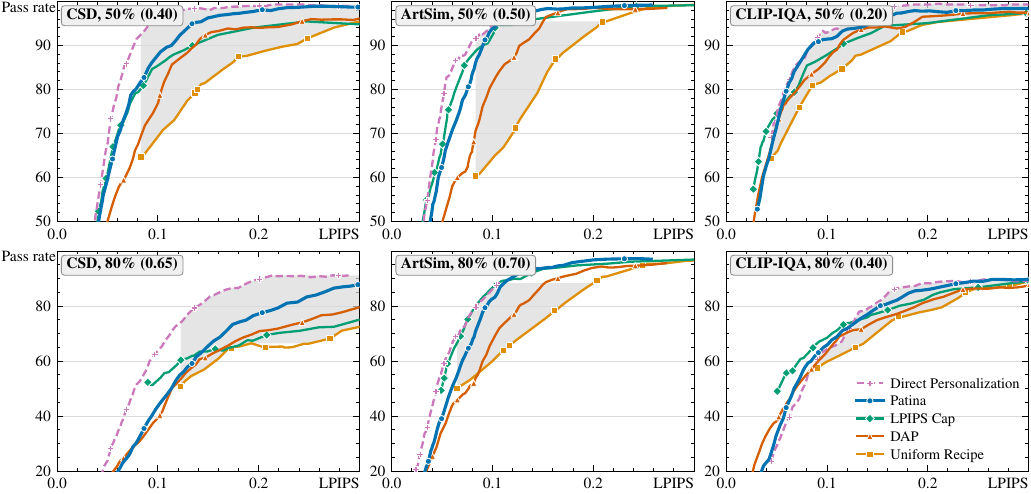}
  \vspace{-20pt}
  \caption{\textbf{Pass rate against mean LPIPS for five strategies and all scores.} Pass rate (\%) of each strategy at the 50\% threshold (top) and the 80\% threshold (bottom), with the value of $\tau$ in parentheses.}
  \label{fig:curves-all}
\end{figure*}

\subsection{Does the Spread Hold in Other Settings?}
\label{app:backbones}

When SD~1.5 or SDXL is personalized on the images protected through SD~2.1, the standard deviation of the protection score across artworks is 0.17 to 0.28 and 1.6 to 2.9 times the standard deviation across seeds (\cref{fig:backbones}).
Fine-tuning the whole denoiser of SD~2.1 gives a standard deviation across artworks of 0.19 to 0.32, as large as or larger than the 0.17 to 0.22 of SD~2.1 LoRA with the same three methods.
Portfolios personalized together differ in protection as much as single artworks do under the same 18 recipes (\cref{tab:sets-spread}), so training on several artworks of one artist does not average the difference away.
The dependence of protection on the artwork therefore holds in every personalization setting we test, from single artworks to portfolios.

\subsection{What Makes an Artwork Hard?}
\label{app:hard}

In our investigation, we find that the difficulty of an artwork depends on several of its properties at once.
We define the ease of an artwork as its mean protection score over all 54 recipes, so a higher ease means an easier artwork.

\smallskip\noindent\textbf{Source.}
The three sources of our benchmark differ in difficulty.
OmniStyle is the hardest source under all three scores, with a mean ease of 0.478 under CSD against 0.567 for WikiArt and 0.535 for StyleBench.
Under CSD, 44\% of the OmniStyle artworks fall into the hardest quarter of the 300 artworks, and under \artsim{} and CLIP-IQA half of them do, about twice the share that chance gives.
We suspect that the kind of image explains part of this gap, since OmniStyle holds more flat designs and modern digital art than WikiArt, whose artworks are all paintings.

\smallskip\noindent\textbf{Texture.}
Artworks with complex texture are hard to protect.
We measure the complexity of the texture by the size of the artwork under JPEG compression.
The artwork is split into patches of $128\times128$ pixels, each patch is compressed at JPEG quality 50, and the size of a uniform gray patch is subtracted from its size.
The mean number of bits per pixel over the patches is larger for an artwork with detailed texture, since JPEG needs more bits to encode fine detail.
When the artworks are split into four groups by this measure, the mean ease falls from the simplest to the most complex quarter by 0.17 for CSD, 0.20 for \artsim{}, and 0.10 for CLIP-IQA.
Splitting each source separately gives nearly the same drop, so texture matters within every source.
The difference between sources, however, does not come from texture.
After we remove the part of the ease that a linear fit on the texture rank explains, OmniStyle still lies 0.065 and 0.097 below the overall mean under CSD and \artsim{}, close to its gaps of 0.069 and 0.102 before this texture correction.

\smallskip\noindent\textbf{Imitation by personalization.}
Hönig \etal~\cite{honig2025adversarial} relate the variation across artists to how well personalization imitates the original artwork, since an artist whose style the model imitates poorly in the first place appears well protected.
Our style scores divide by how much the original personalization learns the style (\cref{app:score}), so poor imitation alone does not make an artwork look well protected.
Splitting the artworks by how much the original personalization learns changes the mean ease between quarters by at most 0.09, in a direction that differs between the scores, and the quarter that personalization imitates least is never the easiest.

No single property therefore separates easy from hard artworks, which is why the \ours{} predictor combines many signals of the artwork with its response to a trial protection (\cref{sec:predict}).
\Cref{fig:qual-easy,fig:qual-medium,fig:qual-hard} show easy, medium, and hard artworks with the recipes that five strategies of \cref{sec:recipe} give them at mean LPIPS 0.15 (\cref{app:perartwork}).

\section{Predicting Protection in Advance}
\label{app:predict}

\begin{table}[t]
  \centering
  \caption{\textbf{Candidate signals.} We keep 63 of 131 candidate signals using only their correlations with one another and group them into the four categories of \cref{sec:signals}.}
  \label{tab:signals}
  \footnotesize
  \begin{tabularx}{\linewidth}{@{}lX@{}}
    \toprule
    Group & Kept Signals \\
    \midrule
    \multicolumn{2}{@{}l}{\emph{Image}} \\
    Frequency & spectrum slope, low-band energy, anisotropy \\
    Texture & JPEG compression complexity, spread of local complexity, pattern repetition \\
    Color & brightness, contrast, colorfulness \\
    VAE & latent entropy, high-frequency latent energy, encoder sensitivity \\
    Base Model & denoising loss at mid noise, gain from the text condition, match to the class prompt \\
    \midrule
    \multicolumn{2}{@{}l}{\emph{Protection Trace}} \\
    Loss Curve & first loss, relative gain, saturation \\
    Input Gradient & mean, decay, share at the bound, distance to the target spectrum, surrogate loss (ACE) \\
    Size & $L_2$ norm, LPIPS \\
    Frequency & mid-band energy, mid-band DCT energy, mid-band phase change \\
    Placement & share in flat regions, correlation with texture and with orientation \\
    \midrule
    \multicolumn{2}{@{}l}{\emph{Protected Image}} \\
    VAE Latent & latent shift, encoder gain, low-band latent shift \\
    VAE Distribution & posterior spread ratio, round-trip mismatch, Gram distance \\
    SDEdit & difficulty estimate of Impasto, style match after SDEdit at 0.2, its drop at 0.4 \\
    Survival & high-pass effect, survival after JPEG compression and after resizing \\
    Model Response & denoising loss change at low and high noise, shift of a style embedding \\
    \midrule
    \multicolumn{2}{@{}l}{\emph{LoRA Probe}} \\
    Gradient Statistics & gradient norm of the LoRA weights on the original, spread of the norms across noise levels, share of the update that passes through the text condition \\
    Answer Contrast & change of the denoiser output on the original after a few LoRA steps, the part of this change that moves toward the artwork, and the part that the perturbation absorbs \\
    Contrast Decomposition & the same two parts estimated from gradients before training, and the first part at high noise \\
    Style & the same two parts measured on style features \\
    Training Dynamics & loss variance, loss drop on the original query, flatness \\
    Band Answer & drift of the output for the class prompt, and the two parts in the mid-frequency band \\
    \bottomrule
  \end{tabularx}
\end{table}

\begin{table}[t]
  \centering
  \caption{\textbf{Calibration for an unseen protection method.} We leave one protection method out of training and calibrate every prediction with a few artworks personalized under it (\cref{app:unseen}). Rows give the number of these artworks, and values are RMSE. With none, the \ours{} predictor is trained on the other five methods and mean report is their average, and All is the predictor trained on the method itself, as in \cref{tab:predict}. IQA is CLIP-IQA.}
  \label{tab:calibration}
  \footnotesize
  \setlength{\tabcolsep}{0pt}
  \begin{tabular*}{\linewidth}{@{\extracolsep{\fill}}lcccccc@{}}
    \toprule
    & \multicolumn{3}{c}{Mean Report} & \multicolumn{3}{c}{\textbf{\ours{} Predictor}} \\
    \cmidrule(lr){2-4}\cmidrule(lr){5-7}
    Artworks & CSD\,$\downarrow$ & \artsim{}\,$\downarrow$ & IQA\,$\downarrow$ & CSD\,$\downarrow$ & \artsim{}\,$\downarrow$ & IQA\,$\downarrow$ \\
    \midrule
    0 & 0.246 & 0.283 & 0.194 & \textbf{0.201} & \textbf{0.233} & \textbf{0.180} \\
    5 & 0.232 & 0.262 & 0.174 & \textbf{0.194} & \textbf{0.225} & \textbf{0.161} \\
    10 & 0.222 & 0.253 & 0.167 & \textbf{0.187} & \textbf{0.217} & \textbf{0.156} \\
    All & 0.213 & 0.241 & 0.160 & \textbf{0.166} & \textbf{0.198} & \textbf{0.139} \\
    \bottomrule
  \end{tabular*}
\end{table}

This section gives the details and further results of \cref{sec:predict}, which answers Q2 of the introduction.

\subsection{Signals and Predictor}
\label{app:signals}

\smallskip\noindent\textbf{Signals.}
\Cref{tab:signals} lists the signals of the four categories of \cref{sec:signals}, grouped by what they measure.
Image signals describe the artwork before protection, through its frequency content, texture, and color and through the response of the VAE and the base model to it.
Protection trace signals record how the protection optimization proceeds, through its loss curve and input gradient and through the size, frequency content, and placement of the perturbation.
Protected image signals measure how the finished protected image acts on the models, through the VAE, a light SDEdit~\cite{meng2022sdedit}, simple transforms that the perturbation must survive, and the denoiser.
LoRA probe signals approximate personalization with a few LoRA steps, and the \ours{} predictor does not use them.
Within each group, we cluster the 131 candidates by their rank correlations with each other and keep one representative per cluster, which leaves 63 signals.
The selection never uses the protection scores of the artworks.
Among predictors that use a single group of signals, the SDEdit group gives the lowest RMSE on every score (CSD 0.174), and removing it from the \ours{} predictor raises the CSD RMSE by 4.0\%.

\smallskip\noindent\textbf{Predictor.}
The predictor is LightGBM with learning rate 0.05, four leaves, 140 trees, row and column subsampling of 0.6, at least 20 samples per leaf, $L_2$ regularization of 1.0, and a monotone increasing constraint on strength.
The folds split the 300 artworks, so all recipes of an artwork stay in one fold, and placing the artworks of each known artist in one fold leaves the results of \ours{} nearly unchanged (CSD pass rate 96.5\% against 96.2\%).
False pass in \cref{tab:predict} is the percentage of the artworks predicted to pass the 50\% threshold whose measured score falls below it, and for direct personalization it equals the rate at which a recipe that passes the threshold under one seed falls below it under another.
The image embedding baseline takes the CLS token of DINOv2~\cite{oquab2024dinov2} with a ViT-L/14 backbone on the original, scales it to unit length, and fits a ridge regression with $\alpha=1$ from the embedding to the scores of all 54 recipes, with the same cross-validation folds as the \ours{} predictor.

\subsection{Beyond the Training Setup}
\label{app:unseen}

\label{app:models}

\begin{table}[t]
  \centering
  \caption{\textbf{Prediction for other personalization settings by method.} We apply the predictor trained with LoRA on SD~2.1 to each method at 16/255 in the settings of \cref{tab:transfer}. Values are RMSE, and both predictions are calibrated with the artworks personalized in the new setting other than the scored ones. DreamBooth fine-tunes the whole denoiser of SD~2.1.}
  \label{tab:models}
  \footnotesize
  \setlength{\tabcolsep}{0pt}
  \begin{tabular*}{\linewidth}{@{\extracolsep{\fill}}llcccc@{}}
    \toprule
    & & \multicolumn{2}{c}{Mean Report} & \multicolumn{2}{c}{\textbf{\ours{} Predictor}} \\
    \cmidrule(lr){3-4}\cmidrule(lr){5-6}
    Setting & Method & CSD\,$\downarrow$ & \artsim{}\,$\downarrow$ & CSD\,$\downarrow$ & \artsim{}\,$\downarrow$ \\
    \midrule
    SD~1.5 & AdvDM & 0.254 & 0.302 & \textbf{0.194} & \textbf{0.261} \\
     & PhotoGuard & 0.231 & 0.163 & \textbf{0.225} & \textbf{0.150} \\
     & StyleGuard & 0.254 & 0.338 & \textbf{0.219} & \textbf{0.295} \\
    SDXL & AdvDM & 0.212 & 0.230 & \textbf{0.135} & \textbf{0.227} \\
     & PhotoGuard & 0.225 & 0.192 & \textbf{0.218} & \textbf{0.181} \\
     & StyleGuard & 0.195 & 0.306 & \textbf{0.143} & \textbf{0.270} \\
    DreamBooth & AdvDM & 0.225 & 0.242 & \textbf{0.155} & \textbf{0.212} \\
     & PhotoGuard & 0.244 & 0.365 & \textbf{0.234} & \textbf{0.356} \\
     & StyleGuard & 0.223 & 0.335 & \textbf{0.167} & \textbf{0.287} \\
    \bottomrule
  \end{tabular*}
\end{table}

\smallskip\noindent\textbf{Unseen sources and protection methods.}
For an unseen source, all artworks of one of the three sources are left out of training, the predictors are trained on the other two, and \cref{tab:general} averages the RMSE over the three sources.
For an unseen protection method, one of the six methods is left out, and every predictor is trained on the other five without the name of the method.
Mean report then gives the average score of the other five methods at each strength, and image embedding predicts this average from the artwork.
Predictions for a method left out of training can be off by the same amount for every artwork at a given strength, so \cref{tab:general} calibrates them with five training artworks protected with the new method and personalized.
At each strength, every prediction is shifted by the gap between its mean over these five artworks and their measured mean, which turns mean report into this measured mean, and the RMSE is averaged over 20 draws of these artworks.

On each source held out of training, the predictor is far more accurate than mean report, for example on OmniStyle (CSD RMSE 0.190 against 0.242).
\Cref{tab:calibration} varies the number of personalized artworks used for calibration.
Without any of them, the \ours{} predictor is already more accurate than mean report, and five or ten artworks bring it closer to the predictor trained on the method.
Calibrating its level with five personalized artworks of the method brings its RMSE below the mean report of that method on the two style scores (0.194 and 0.225 against 0.213 and 0.241).

\begin{figure}[t]
  \centering
  \includegraphics[width=\linewidth]{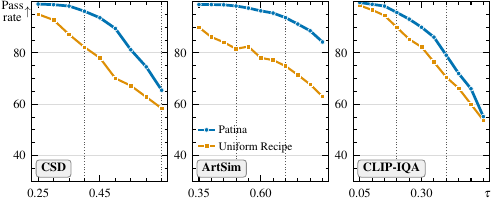}
  \vspace{-20pt}
  \caption{\textbf{Pass rate (\%) as the threshold varies.} We vary the threshold $\tau$ of each score and compare uniform recipe with \ours{} at a mean LPIPS limit of 0.15. Dotted lines mark the 50\% and the 80\% thresholds of each score.}
  \label{fig:sweep}
\end{figure}

\smallskip\noindent\textbf{Other personalization settings.}
The predictor trained on SD~2.1 LoRA is applied without retraining to the runs with SD~1.5, SDXL, and DreamBooth on SD~2.1 (\cref{tab:models}).
The overall level of protection differs between settings, so a protection tool calibrates the predictions once in each setting with artworks personalized there, and mean report becomes the mean score of the same artworks.
The scored artworks are left out of this calibration, with the folds of the main evaluation.
With this calibration, the predictor is more accurate than mean report for every setting and method on both style scores, so the differences between artworks that it learns on SD~2.1 LoRA transfer to the other settings.
For few-shot personalization, the predictor trained on all 300 single artworks predicts each artwork of a portfolio from its own trial protection without retraining, and the mean over the artworks gives the prediction for the portfolio.
Each prediction, including mean report, is calibrated with the other 19 portfolios, and the mean prediction of \ours{} is more accurate than mean report, DAP, and image embedding on both style scores (\cref{tab:sets}), as on single artworks.

\section{Choosing Protection per Artwork}
\label{app:recipe}

This section gives the details and further results of \cref{sec:recipe}, which answers Q3 of the introduction, from the cost of each strategy and the parts of \ours{} to its visibility and its protection under purification and for portfolios.

\setcounter{topnumber}{1}

\begin{table*}[t]
  \centering
  \caption{\textbf{LPIPS cap with parts of \ours{}.} Predicted LPIPS cap sets the strength from the LPIPS that the signals of \ours{} predict, and LPIPS cap + method choice gives each artwork the method that \ours{} predicts to protect best within the limit. The columns and blocks follow \cref{tab:main}, and bold marks the best of the four strategies.}
  \label{tab:lpips-family}
  \footnotesize
  \setlength{\tabcolsep}{0pt}
  \begin{tabular*}{\linewidth}{@{\extracolsep{\fill}}lcccccccccc@{}}
    \toprule
    \multirow{2}{*}[-2.5pt]{Strategy} & \multirow{2}{*}[-2.5pt]{Cost} & \multicolumn{3}{c}{$s<\tau$, All Artworks (\%)} & \multicolumn{3}{c}{$s<\tau$, Hardest 25\% (\%)} & \multicolumn{3}{c}{LPIPS for a 90\% Pass Rate} \\
    \cmidrule(lr){3-5}\cmidrule(lr){6-8}\cmidrule(lr){9-11}
    & & CSD\,$\downarrow$ & \artsim{}\,$\downarrow$ & CLIP-IQA\,$\downarrow$ & CSD\,$\downarrow$ & \artsim{}\,$\downarrow$ & CLIP-IQA\,$\downarrow$ & CSD\,$\downarrow$ & \artsim{}\,$\downarrow$ & CLIP-IQA\,$\downarrow$ \\
    \midrule
    \multicolumn{11}{l}{\emph{50\% threshold ($\tau$) with a mean LPIPS limit of 0.15}} \\
    LPIPS Cap & 0.67--2.67 & 7.6 & 3.7 & 6.1 & 9.9 & 7.8 & 11.2 & 0.135 & 0.090 & 0.113 \\
    Predicted LPIPS Cap & 0.28 & 8.4 & 3.7 & 6.2 & 11.8 & 9.1 & 13.2 & 0.117 & \textbf{0.084} & 0.121 \\
    LPIPS Cap + Method Choice & 0.95--2.95 & 6.5 & 5.1 & 5.1 & 12.7 & 9.7 & \textbf{7.4} & 0.112 & 0.095 & 0.097 \\
    \textbf{\ours{}} & 0.28 & \textbf{3.8} & \textbf{1.7} & \textbf{4.3} & \textbf{3.0} & \textbf{4.7} & 11.1 & \textbf{0.110} & 0.090 & \textbf{0.087} \\
    \midrule
    \multicolumn{11}{l}{\emph{80\% threshold ($\tau$) with a mean LPIPS limit of 0.20}} \\
    LPIPS Cap & 0.67--2.67 & 31.9 & 4.8 & 16.9 & 53.5 & 11.6 & 32.8 & -- & 0.123 & -- \\
    Predicted LPIPS Cap & 0.28 & 32.2 & 5.1 & 17.5 & 54.9 & 13.7 & 31.9 & -- & \textbf{0.116} & -- \\
    LPIPS Cap + Method Choice & 0.95--2.95 & 25.2 & 6.2 & 16.0 & 38.9 & 10.9 & 34.8 & 0.306 & 0.147 & 0.296 \\
    \textbf{\ours{}} & 0.28 & \textbf{22.8} & \textbf{3.3} & \textbf{13.9} & \textbf{33.1} & \textbf{9.4} & \textbf{27.4} & \textbf{0.275} & 0.118 & \textbf{0.259} \\
    \bottomrule
  \end{tabular*}
\end{table*}

\begin{table}[!t]
  \centering
  \caption{\textbf{Pass rate at other operating points.} We compare the strategies at an equal mean $L_\infty$ strength (/255) and at a mean LPIPS limit of 0.10, under the 50\% threshold.}
  \label{tab:operating}
  \footnotesize
  \setlength{\tabcolsep}{0pt}
  \begin{tabular*}{\linewidth}{@{\extracolsep{\fill}}lcccc@{}}
    \toprule
    & \multicolumn{3}{c}{Mean $L_\infty$ (/255)} & Mean LPIPS \\
    \cmidrule(lr){2-4}\cmidrule(lr){5-5}
    Pass rate (\%)\,$\uparrow$ & 6 & 8 & 10 & 0.10 \\
    \midrule
    \multicolumn{5}{l}{\emph{CSD}} \\
    Uniform Recipe & 80.1 & 87.7 & 90.1 & 69.4 \\
    DAP~\cite{ahn2024impasto} & 81.6 & 92.4 & 94.0 & 77.2 \\
    LPIPS Cap & 75.8 & 87.7 & 93.0 & 85.7 \\
    \textbf{\ours{}} & \textbf{87.6} & \textbf{95.5} & \textbf{97.4} & \textbf{87.7} \\
    \midrule
    \multicolumn{5}{l}{\emph{\artsim{}}} \\
    Uniform Recipe & 71.3 & 86.9 & 95.3 & 64.5 \\
    DAP~\cite{ahn2024impasto} & 79.2 & 91.1 & 98.2 & 81.7 \\
    LPIPS Cap & 83.1 & 94.7 & 96.6 & 92.3 \\
    \textbf{\ours{}} & \textbf{87.8} & \textbf{97.5} & \textbf{98.4} & \textbf{93.6} \\
    \midrule
    \multicolumn{5}{l}{\emph{CLIP-IQA}} \\
    Uniform Recipe & 81.0 & 87.6 & 93.0 & 82.6 \\
    DAP~\cite{ahn2024impasto} & 79.9 & 88.8 & 94.0 & 87.0 \\
    LPIPS Cap & 77.4 & 87.4 & 92.1 & 88.0 \\
    \textbf{\ours{}} & \textbf{88.5} & \textbf{92.8} & \textbf{95.7} & \textbf{91.6} \\
    \bottomrule
  \end{tabular*}
\end{table}

\begin{table}[t]
  \centering
  \caption{\textbf{Choosing with one score and evaluating with another.} Each strategy chooses recipes with the prediction of one score, and we evaluate them with each of the three scores, as the pass rate (\%) at a mean LPIPS limit of 0.15 under the 50\% threshold.}
  \label{tab:cross}
  \footnotesize
  \setlength{\tabcolsep}{0pt}
  \begin{tabular*}{\linewidth}{@{\extracolsep{\fill}}llccc@{}}
    \toprule
    & & \multicolumn{3}{c}{Evaluated With} \\
    \cmidrule{3-5}
    Chosen With & Strategy & CSD\,$\uparrow$ & \artsim{}\,$\uparrow$ & CLIP-IQA\,$\uparrow$ \\
    \midrule
    CSD & Uniform Recipe & 82.2 & 77.8 & 72.2 \\
     & \textbf{\ours{}} & 96.2 & 88.9 & 74.9 \\
    \artsim{} & Uniform Recipe & 79.2 & 81.6 & 64.9 \\
     & \textbf{\ours{}} & 89.0 & 98.3 & 76.1 \\
    CLIP-IQA & Uniform Recipe & 48.5 & 84.0 & 89.9 \\
     & \textbf{\ours{}} & 56.3 & 87.6 & 95.7 \\
    \bottomrule
  \end{tabular*}
\end{table}

\begin{table*}[t]
  \centering
  \caption{\textbf{Analysis of \ours{} with visibility.} We extend \cref{tab:ablation} with the mean LPIPS at which the pass rate reaches 90\% under the 50\% threshold. Rate is the pass rate (\%) at a mean LPIPS limit of 0.15, and the rows follow \cref{tab:ablation}. The last row leaves each method in turn out of the recipe set and averages the results over the six methods.}
  \label{tab:ablation-full}
  \footnotesize
  \setlength{\tabcolsep}{0pt}
  \begin{tabular*}{\linewidth}{@{\extracolsep{\fill}}lcccccc@{}}
    \toprule
    & \multicolumn{2}{c}{CSD} & \multicolumn{2}{c}{\artsim{}} & \multicolumn{2}{c}{CLIP-IQA} \\
    \cmidrule(lr){2-3}\cmidrule(lr){4-5}\cmidrule(lr){6-7}
    Variant & Rate\,$\uparrow$ & LPIPS\,$\downarrow$ & Rate\,$\uparrow$ & LPIPS\,$\downarrow$ & Rate\,$\uparrow$ & LPIPS\,$\downarrow$ \\
    \midrule
    Uniform Recipe & 82.2 & 0.225 & 81.6 & 0.179 & 89.9 & 0.156 \\
    Image Signals & 92.8 & 0.128 & 97.1 & 0.094 & 95.6 & 0.097 \\
    + Trial Protection Trace & 92.4 & 0.125 & 97.3 & 0.088 & 96.3 & 0.099 \\
    + Trial Protected Image (\textbf{\ours{}}) & 96.2 & 0.110 & 98.3 & 0.090 & 95.7 & 0.087 \\
    \midrule
    Trial Length (Steps) = 10 & 95.0 & 0.113 & 98.2 & 0.095 & 96.1 & 0.087 \\
    Trial Length (Steps) = 25 & 95.3 & 0.111 & 98.3 & 0.090 & 96.5 & 0.088 \\
    Trial Length (Steps) = 50 (\textbf{\ours{}}) & 96.2 & 0.110 & 98.3 & 0.090 & 95.7 & 0.087 \\
    Trial Length (Steps) = 100 & 94.7 & 0.108 & 98.3 & 0.091 & 95.8 & 0.091 \\
    \midrule
    \textbf{\ours{}}, Same Method & 88.6 & 0.166 & 98.0 & 0.091 & 95.4 & 0.090 \\
    \textbf{\ours{}}, All 54 Recipes Protected & 94.9 & 0.099 & 98.5 & 0.073 & 97.0 & 0.088 \\
    \midrule
    \textbf{\ours{}}, One Method Left Out of the Recipe Set & 95.4 & 0.112 & 97.2 & 0.097 & 93.8 & 0.105 \\
    \bottomrule
  \end{tabular*}
\end{table*}

\begin{table*}[t]
  \centering
  \caption{\textbf{Per-artwork visibility.} We compare the per-artwork LPIPS at a mean LPIPS of 0.15 under CSD, and the percentage of artworks below the 50\% threshold ($s<\tau$) when every strategy gets the same per-artwork LPIPS limit and moves its knob to a mean LPIPS of 0.15.}
  \label{tab:perartwork}
  \footnotesize
  \setlength{\tabcolsep}{0pt}
  \begin{tabular*}{\linewidth}{@{\extracolsep{\fill}}lcccccccc@{}}
    \toprule
    & \multicolumn{2}{c}{Per-Artwork LPIPS (CSD)} & \multicolumn{3}{c}{$s<\tau$ (\%), Limit 0.20} & \multicolumn{3}{c}{$s<\tau$ (\%), Limit 0.25} \\
    \cmidrule(lr){2-3}\cmidrule(lr){4-6}\cmidrule(lr){7-9}
    Strategy & 90th Percentile\,$\downarrow$ & Maximum\,$\downarrow$ & CSD\,$\downarrow$ & \artsim{}\,$\downarrow$ & CLIP-IQA\,$\downarrow$ & CSD\,$\downarrow$ & \artsim{}\,$\downarrow$ & CLIP-IQA\,$\downarrow$ \\
    \midrule
    Uniform Recipe & 0.275 & 0.468 & 11.0 & 3.3 & 7.1 & 13.7 & 7.8 & 8.8 \\
    Image Embedding & 0.262 & 0.464 & 5.5 & \textbf{2.8} & 7.2 & 5.3 & 2.1 & 6.1 \\
    DAP~\cite{ahn2024impasto} & 0.228 & 0.332 & 7.9 & 2.9 & 7.1 & 7.4 & 2.4 & 6.9 \\
    LPIPS Cap & \textbf{0.184} & \textbf{0.200} & 7.9 & 3.2 & 6.7 & 8.5 & 3.6 & 6.1 \\
    \textbf{\ours{}} & 0.237 & 0.394 & \textbf{4.2} & 2.9 & \textbf{5.8} & \textbf{3.7} & \textbf{1.7} & \textbf{4.5} \\
    \bottomrule
  \end{tabular*}
\end{table*}

\subsection{Details of the Strategy Comparison}
\label{app:main}

\smallskip\noindent\textbf{Strategies and cost.}
Uniform recipe gives every artwork the weakest recipe whose pass rate on the training artworks of each fold reaches a target, with the best method at that strength.
Per-source recipe and per-cluster recipe give one recipe to each source or to each of 30 style clusters, which per-cluster recipe finds by $k$-means on the CSD embeddings of the artworks.
LPIPS cap needs the LPIPS of each artwork at each strength, so it protects the artwork several times.
One protection of 100 steps costs about a third of a personalization, and costs exclude the final protection, which every strategy needs. The extra protections of a binary search over strengths then add 0.67 to 1.0 personalizations, and protecting at all nine strengths adds about 2.67.
Image embedding and DAP predict the scores from the DINOv2 embedding of the artwork and from the difficulty estimate of Impasto on the trial protection of \ours{}, respectively, so image embedding needs only the embedding and DAP runs the same trial protection as \ours{}.

\ours{} with all 54 recipes protected applies all 54 recipes to the artwork and measures the signals of each protected image, at a cost of about 22 personalizations.
Direct personalization protects the artwork with all 54 recipes, personalizes each protected image with one seed, chooses with these measured scores, and is scored on the other two seeds.
It needs the same 54 protections without the signals, about 18 personalizations, and then personalizes each protected image once, so it costs about 71.7 personalizations without the final protection.
On \artsim{} at mean LPIPS 0.15, \ours{} exceeds direct personalization, which chooses from the scores of a single seed and inherits the noise of that seed.
The hardest quarter in \cref{tab:main} consists of the 25 artworks that need the strongest recipe under the mean of three seeds.

\smallskip\noindent\textbf{LPIPS cap with parts of \ours{}.}
To see whether the gain of \ours{} comes from one of its parts alone, we give LPIPS cap two of them (\cref{tab:lpips-family}).
Predicted LPIPS cap sets the strength from the LPIPS that the signals of \ours{} predict, so it needs no extra protection.
LPIPS cap + method choice gives each artwork the method that \ours{} predicts to protect best within the limit, which adds the trial protection and the signals of \ours{} to the cost of LPIPS cap.
\ours{} still leaves fewer artworks below the threshold than both variants in almost every column, so its gain does not come from predicting the visibility or choosing the method alone.

\smallskip\noindent\textbf{Strength by difficulty.}
If the method were chosen for each artwork in hindsight from the measured scores, more than three quarters of the artworks would pass the protection threshold at 4/255 or less (CSD 76\%, \artsim{} 80\%, CLIP-IQA 81\%).
Uniform recipe gives the easier and the harder half of the artworks the same strength, at which the perturbation is more visible on the easier half (CSD LPIPS 0.207 against 0.093), and \ours{} moves strength from the easier artworks to the harder ones (\cref{fig:q3}c,d).
In the five groups of \cref{fig:q3}c,d, \ours{} keeps 96\% of the hardest fifth above the protection threshold under CSD, where uniform recipe keeps 43\%, and lowers the LPIPS of the easiest fifth from 0.244 to 0.166.
LPIPS cap gives every artwork the same LPIPS limit, which leaves 5.6\% of the easiest quarter below the protection threshold under CSD, while \ours{} keeps 98 to 100\% of these artworks above it.

\subsection{Robustness across Thresholds and Scores}
\label{app:choice}
\label{app:threshold}

\Cref{fig:q3} shows CSD and \artsim{} at the 50\% threshold, and \cref{fig:curves-all} adds DAP and LPIPS cap, CLIP-IQA, and the 80\% threshold, where \ours{} stays above uniform recipe on every score.
The gain of \ours{} over uniform recipe holds at every threshold of every score, which includes the 50\% and the 80\% thresholds (\cref{fig:sweep}).
At an equal pass rate of 90\%, \ours{} reaches a higher visibility compared to the other baselines (\cref{tab:visibility}).
LPIPS cap chooses by LPIPS, so the comparison at equal mean LPIPS favors it.
At equal mean $L_\infty$ strength, \ours{} lets the most artworks pass the protection threshold, while LPIPS cap, which spreads the visibility evenly, stays close to or below uniform recipe at 6 and 8/255 under CSD and CLIP-IQA (\cref{tab:operating}).
Choosing with the prediction of one score, \ours{} also lets more artworks pass the protection threshold than uniform recipe when evaluated with the other two scores (\cref{tab:cross}).

\subsection{\texorpdfstring{Which Parts of \ours{} Matter?}{Which Parts of Patina Matter?}}
\label{app:trial}

\Cref{tab:ablation-full} extends \cref{tab:ablation} with the mean LPIPS at which each variant reaches a pass rate of 90\%.
Under CSD, the signals lower this LPIPS from 0.225 under uniform recipe to 0.110 under \ours{}, trials of 10 to 100 steps stay within 0.003 of it, and the same method for all artworks needs a 51\% higher LPIPS.
Any recipe also works as the trial.
With each of the 54 recipes as a full-length trial protection, the pass rate at mean LPIPS 0.15 ranges from 92.5 to 96.5\% for CSD, 96.0 to 98.6\% for \artsim{}, and 95.6 to 97.1\% for CLIP-IQA, all above uniform recipe.

\setcounter{topnumber}{2}
\subsection{How Visible Is Each Protected Artwork?}
\label{app:perartwork}

\smallskip\noindent\textbf{Per-artwork limit.}
\ours{} gives the hard artworks more strength, so an artist may want to bound the visibility of every artwork.
At mean LPIPS 0.15, the 90th percentile of the per-artwork LPIPS under CSD is 0.24 for \ours{}, against 0.28 for uniform recipe and 0.18 for LPIPS cap, which bounds every artwork by design (\cref{tab:perartwork}).
We therefore give every strategy the same per-artwork limit, so that the strength steps down within the same method whenever the measured LPIPS of the chosen recipe exceeds the limit, and each strategy moves its knob again to a mean LPIPS of 0.15.
Under this condition, \ours{} leaves the fewest artworks below the threshold under CSD and CLIP-IQA at a limit of 0.20 and under all three scores at 0.25.
A limit also helps uniform recipe, since stepping down the few artworks that are too visible lets its knob protect the other artworks more strongly at the same mean LPIPS.

\smallskip\noindent\textbf{Qualitative examples.}
\Cref{fig:qual-compare} shows six validation artworks with the recipe that uniform recipe and \ours{} give each of them at mean LPIPS 0.15 (CSD).
Uniform recipe is chosen on the training artworks of each cross-validation fold, so the six artworks come from one fold, the largest fold whose choice is the most common one (AdvDM at 6/255), and they all receive this recipe.
We split this fold into an easier and a harder half by the mean CSD score of each artwork over the 54 recipes.
The easier three show how \ours{} lowers the LPIPS, and the harder three how a stronger recipe lets an artwork pass.
Each generated image is the first one generated after personalization, labeled with the CSD score averaged over three seeds.

\begin{figure*}[t]
  \centering
  \includegraphics[width=\linewidth]{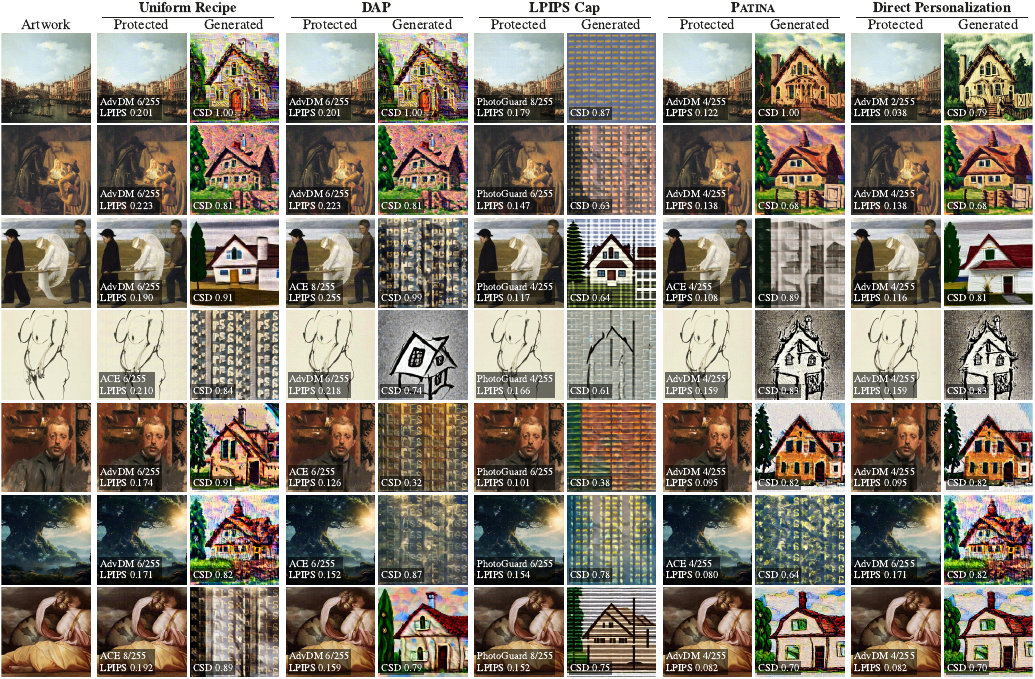}
  \caption{\textbf{Recipes for easy artworks.} Seven validation artworks that \ours{} protects at 4/255, with the recipe that each strategy gives them at mean LPIPS 0.15. Each protected image shows its recipe and LPIPS, and each generated image its CSD protection score averaged over three seeds. All generated images use the prompt ``an image of a house in $S^*$ style''.}
  \label{fig:qual-easy}
\end{figure*}

\begin{figure*}[t]
  \centering
  \includegraphics[width=\linewidth]{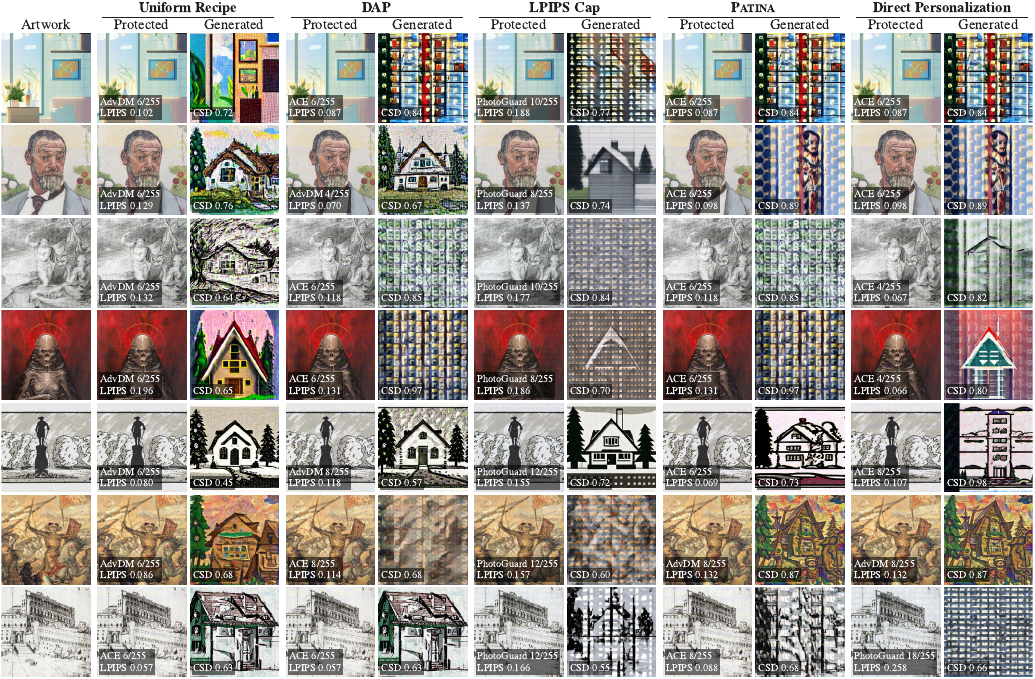}
  \caption{\textbf{Recipes for medium artworks.} Seven validation artworks that \ours{} protects at 6/255 to 10/255, shown as in \cref{fig:qual-easy}.}
  \label{fig:qual-medium}
\end{figure*}

\begin{figure*}[t]
  \centering
  \includegraphics[width=\linewidth]{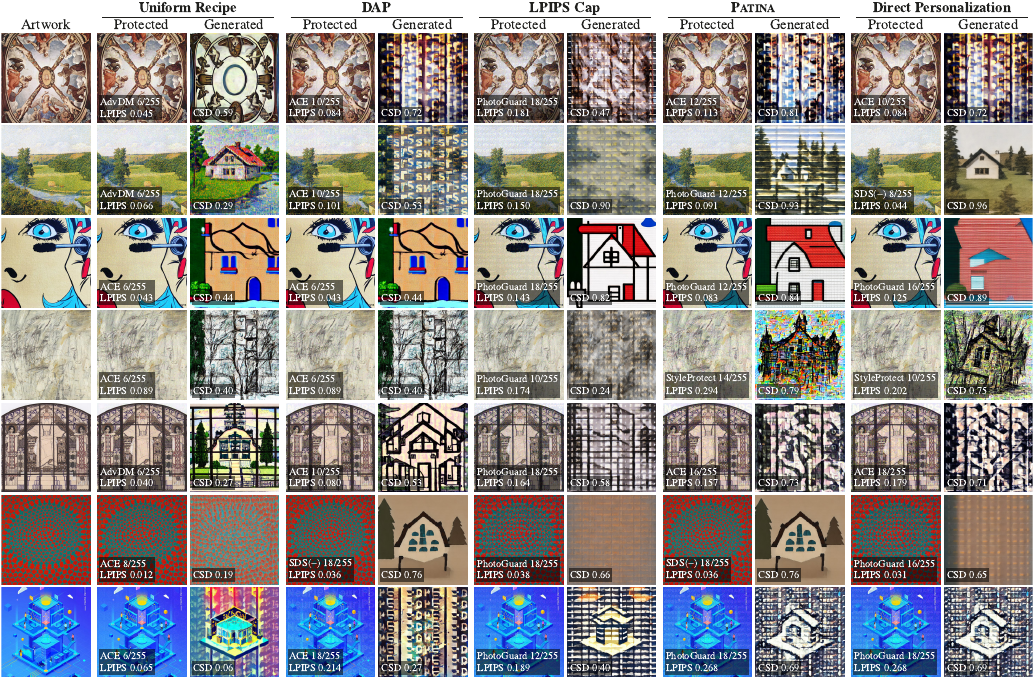}
  \caption{\textbf{Recipes for hard artworks.} Seven validation artworks that \ours{} protects at 12/255 to 18/255, shown as in \cref{fig:qual-easy}.}
  \label{fig:qual-hard}
\end{figure*}

\Cref{fig:qual-easy,fig:qual-medium,fig:qual-hard} extend \cref{fig:qual-compare} to five strategies of \cref{tab:main} with the same labels.
They group the validation artworks by the strength of the recipe that \ours{} gives them, 4/255 for easy, 6/255 to 10/255 for medium, and 12/255 to 18/255 for hard artworks, and show seven artworks of each group, two or three of them from sources other than WikiArt.
Uniform recipe, LPIPS cap, and the predictors are those of the cross-validation fold that holds each artwork out, and direct personalization chooses with single-seed scores.

\subsection{Unseen Sources and Methods}
\label{app:unseen-choice}

With one source held out of training, uniform recipe reaches a pass rate of 63.6\% on StyleBench and 70.3\% on OmniStyle under CSD, while \ours{} reaches 96.2\% and 88.3\%, so measuring each artwork compensates for the shift between sources.
The last row of \cref{tab:ablation-full} treats each protection method in turn as missing from the recipe set.
Leaving one method out lowers the pass rate only slightly, because the other methods substitute for it through the shared structure.
Training the predictor on the scores of the method raises the pass rate on all three scores, so a new method can join the recipe set once it is scored on the training artworks.

\subsection{Protection under Purification}
\label{app:purify}

These runs use 100 artworks and three methods.
The weakest of their strengths, 6/255, already has a mean LPIPS of 0.14, so every strategy sits at the low end of its curve near 0.15, and \cref{tab:purify-app} compares at a mean LPIPS limit of 0.20.
Under JPEG, a choice made for clean personalization gives easy artworks weak perturbations, which JPEG removes first.
\ours{} chosen for clean personalization therefore falls below uniform recipe at low LPIPS limits on the two style scores (\cref{tab:jpeg}).
The purification-aware mode targets the score after JPEG and lets more artworks pass the threshold than uniform recipe with the same target at every limit it reaches.
After Noisy Upscaling, which removes most of the protection, the useful question for an artwork becomes whether any recipe is predicted to pass the threshold, and the warning of \ours{} answers it.
\Cref{tab:purify-app} gives each artwork under Noisy Upscaling the method that \ours{} predicts to protect best at 18/255, and the artworks with the warning pass the threshold far less often than the other artworks.

\subsection{Few-Shot Personalization}
\label{app:sets}

\begin{table}[!t]
  \centering
  \caption{\textbf{JPEG with and without purification awareness.} We compare recipes chosen for personalization without purification (Clean) with recipes chosen for the score after JPEG (Aware), as the pass rate (\%) after JPEG at three mean LPIPS limits. Bold marks the better purification-aware strategy in each column, and a dash indicates a strategy that does not reach the limit.}
  \label{tab:jpeg}
  \footnotesize
  \setlength{\tabcolsep}{0pt}
  \begin{tabular*}{\linewidth}{@{\extracolsep{\fill}}llccc@{}}
    \toprule
    & & \multicolumn{3}{c}{Mean LPIPS Limit} \\
    \cmidrule(lr){3-5}
    \multicolumn{2}{@{}l}{Pass rate (\%)\,$\uparrow$} & 0.15 & 0.20 & 0.30 \\
    \midrule
    CSD & Uniform Recipe, Clean & 64.3 & 78.4 & 79.0 \\
     & \textbf{\ours{}}, Clean & 54.6 & 72.6 & 87.2 \\
     & Uniform Recipe, Aware & 60.9 & 79.0 & 89.6 \\
     & \textbf{\ours{}}, Aware & \textbf{63.9} & \textbf{82.4} & \textbf{92.0} \\
    \midrule
    \artsim{} & Uniform Recipe, Clean & 37.9 & 52.7 & 76.8 \\
     & \textbf{\ours{}}, Clean & 33.4 & 52.6 & 78.2 \\
     & Uniform Recipe, Aware & 57.2 & 67.4 & 80.6 \\
     & \textbf{\ours{}}, Aware & \textbf{58.5} & \textbf{74.4} & \textbf{90.7} \\
    \midrule
    CLIP-IQA & Uniform Recipe, Clean & 42.1 & 63.9 & 80.2 \\
     & \textbf{\ours{}}, Clean & 44.7 & 62.2 & 78.4 \\
     & Uniform Recipe, Aware & -- & 62.5 & 82.0 \\
     & \textbf{\ours{}}, Aware & -- & \textbf{66.7} & \textbf{85.3} \\
    \bottomrule
  \end{tabular*}
\end{table}

\ours{} gives each artwork of a portfolio its own recipe by \cref{eq:choose}, with the predictor of \cref{app:models} and ACE and AdvDM as the candidate methods.
For each of the two style scores, the selection level $\gamma$ is set so that the mean LPIPS over the 98 artworks is 0.10, 0.15, 0.20, or 0.25, and each resulting combination of recipes is personalized as a new portfolio, since the artworks of a portfolio are trained together.
The other strategies give one recipe to each portfolio and read its score from the runs with the 18 recipes.
Uniform recipe gives every portfolio the recipe that it chooses on the 300 single artworks, and LPIPS cap, with one method and one limit, takes the strongest strength at which every artwork of the portfolio stays within the limit.
DAP and image embedding average their predictions over the portfolio and choose one recipe with the rule of \ours{}.
In \cref{tab:sets}, each percentage interpolates linearly between the two settings on either side of the LPIPS limit.
\ours{} leaves the fewest portfolios below the threshold on both style scores, so choosing per artwork also works for few-shot personalization.

\end{document}